\documentclass[
 superscriptaddress,
 amsmath,amssymb,
 aps,
 prb,
 twocolumn,
 longbibliography 
]{revtex4-2}

\usepackage{graphicx}
\usepackage{dcolumn}
\usepackage{bm}
\usepackage{xcolor}
\usepackage{tikz}
\usepackage{xfrac}
\usepackage{blindtext}
\usepackage{times}
\usepackage{multirow}
\usepackage{chemformula}
\usepackage{braket}
\usepackage{mathtools}

\usepackage{color}
\usepackage[colorlinks=true, urlcolor=blue, linkcolor=blue, citecolor=blue, pdftex]{hyperref}

\usepackage[normalem]{ulem}

\definecolor{darkblue}{HTML}{004D6B}
\definecolor{darkred}{HTML}{8c1515}
\definecolor{darkgreen}{HTML}{006400}
\usepackage{hyperref}
\hypersetup{
	colorlinks=true,
	urlcolor=darkred,
	citecolor=darkblue,
	linkcolor=darkred,
	breaklinks
}

\usepackage[caption=false]{subfig}

\graphicspath{{figures/}}

\newcommand{\be}{\begin{equation}}
\newcommand{\ee}{\end{equation}}

\newcommand{\bea}{\begin{eqnarray}}
\newcommand{\eea}{\end{eqnarray}}
\newcommand{\beal}{\begin{align}}
\newcommand{\eeal}{\end{align}}

\renewcommand{\vec}[1]{\ensuremath{\mathbf{#1}}}

\newcommand{\abs}[1]{\left| #1 \right|} 
\newcommand{\avg}[1]{\left< #1 \right>} 

\begin{document}

\title{Revealing Quadrupolar Excitations with Non-Linear Spectroscopy} 

\author{Yoshito Watanabe}
\affiliation{Institute for Theoretical Physics, University of Cologne, 50937 Cologne, Germany}

\author{Simon Trebst}
\affiliation{Institute for Theoretical Physics, University of Cologne, 50937 Cologne, Germany}

\author{Ciar\'an Hickey}
\affiliation{School of Physics, University College Dublin, Belfield, Dublin 4, Ireland}
\affiliation{Centre for Quantum Engineering, Science, and Technology, University College Dublin, Dublin 4, Ireland}

\begin{abstract}
Local moments with a spin $S>1/2$ can exhibit a rich variety of elementary quasiparticle excitations, 
such as quadrupolar excitations, that go beyond the dipolar magnons of conventional spin-$1/2$ systems. 
However, the experimental observation of such quadrupolar excitations is often challenging due to the dipolar selection rules of many linear response probes, rendering them invisible. 
Here we show that non-linear spectroscopy, in the form of two-dimensional coherent spectroscopy (2DCS),
can be used to reveal quadrupolar excitations. 
Considering a family of spin-1 Heisenberg ferromagnets with single-ion easy-axis anisotropy as an example, 
we explicitly calculate their 2DCS signature by combining exact diagonalization and generalized spin wave theory.
We further demonstrate that 2DCS can provide access to the quadrupolar weight of an excitation, analogous to how linear response provides access to the dipolar weight. 
Our work highlights the potential of non-linear spectroscopy as a powerful tool to diagnose multipolar excitations in quantum magnets.
\end{abstract}

\maketitle


Quantum magnets admit a veritable zoo of distinct quasiparticle excitations, offering a versatile playground for the investigation of a diverse array of physics and phenomena. Alongside conventional dipolar excitations, such as the well-known magnons of spin-$1/2$ magnets, there is growing interest in the study of elementary excitations with \emph{multipolar} character~\cite{Penc2012, Bai2021, Remund2022}. Such excitations arise in higher-spin magnets, or those with a more complex doublet structure, and greatly enrich the landscape of quasiparticle physics. As an example, in spin-1 systems, a purely on-site quadrupolar $\abs{\Delta M^z} = 2$ excitation is possible as an elementary excitation, usually referred to as a single-ion bound state (SIBS)~\cite{Silberglitt1970}, whereas in spin-1/2 systems, a $\abs{\Delta M^z} = 2$ excitation will necessarily only appear as a composite excitation, consisting of two dipolar excitations. Understanding the interplay of such excitations can provide deeper insights into fundamental phenomena like quasiparticle decay and renormalization~\cite{Zhitomirsky2013}.

Experimentally, studying multipolar excitations is hampered by the fact that they cannot be straightforwardly probed via linear response, which typically exhibits dipolar selection rules. However, there has been some recent progress~\cite{Legros2021, Bai2023} in revealing quadrupolar excitations under certain conditions: Although purely quadrupolar excitations cannot be created by a dipole operator, weakly breaking spin-rotational symmetry can hybridize quadrupolar and dipolar excitations, allowing both to appear even in conventional probes that follow dipolar selection rules. An alternative approach, which does not rely on such hybridization, involves using a probe that can couple with quadrupolar operators via local higher-order processes. This latter approach has been demonstrated with RIXS spectroscopy~\cite{Nag2022}. 

Here, we show that one can straightforwardly access multipolar excitations via \emph{non-linear} response, encoded in a system's non-linear susceptibilities and expressed in terms of higher-order dynamical correlation functions. As an example, for a spin-1 magnet with local moment at site $i$ in the $\ket{+1}$ state, the third-order response can naturally encode elementary quadrupolar excitations via the matrix element $\bra{+1} S_i^+ \ket{0}\bra{0} S_i^+ \ket{-1}\bra{-1} S_i^- \ket{0}\bra{0} S_i^- \ket{+1}$. In general, the $n$-th order response involves $n+1$ spin operators, enabling the observation of $\Delta M^z=(n+1)/2$ excitations for $n$ odd.

\begin{figure}[b]
  \centering
  \includegraphics[width=0.98\columnwidth]{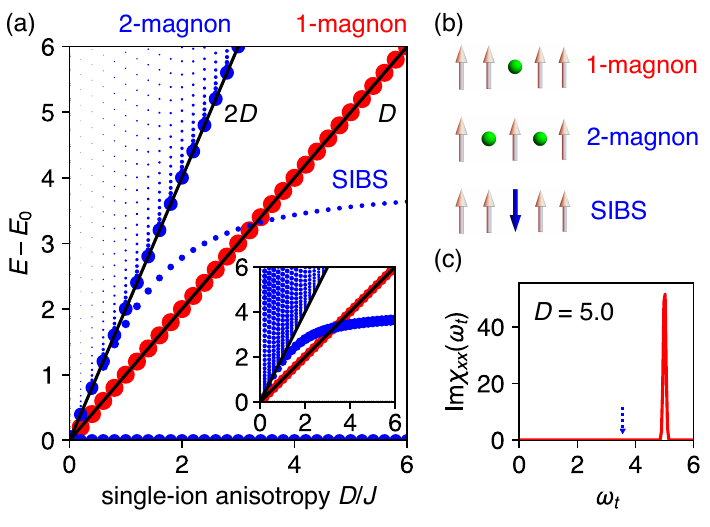}
  \caption{{\bf Invisibility of quadrupolar excitation in linear response}. 
  (a) The energy spectrum, at zero momentum, of the spin-1 FM Heisenberg model with single-ion anisotropy $D$  
	exhibits a 1-magnon excitation, as well as a continuum of 2-magnon excitations and a sharp single-ion bound state (SIBS). 
	The size of the circles is proportional to the transition amplitudes $\abs{\braket{n|M^x|0}}^2$ (red) and $\abs{\braket{n|(M^x)^2|0}}^2$ (blue). 
	Blue circles in the inset indicate $|\braket{n|Q^{x^2-y^2}|0}|^2$. As a representative example, the spectrum shown here is for a 1D $L=100$ site chain. (b) Illustration of the schematic form of the various excitations.
	(c) Linear response at large $D = 5.0$ reveals only the 1-magnon excitation, while the quadrupolar SIBS is completely invisible. 
  }
  \label{fig:chi1_Ddep}
\end{figure}

An ideal technique for probing the nonlinear response of quantum magnets is two-dimensional coherent spectroscopy (2DCS). Using two THz field pulses separated by a time delay $\tau$, one measures the magnetization a measurement time $t$ after the second pulse. The technique can extract the second-order susceptibility $\chi^{(2)}(\omega_t, \omega_\tau)$, as well as two types of third-order susceptibilities: $\chi^{(3;1)}(\omega_t, \omega_\tau)$ and $\chi^{(3;2)}(\omega_t, \omega_\tau)$. There is a significant amount of ongoing research, from both experiment and theory, in trying to better understand THz 2DCS in magnetic materials~\cite{Lu2017, Wan2019, Parameswaran2020, Choi2020, Li2021, Fava2021, Nandkishore2021, Negahdari2023, Hart2023, Fava2023, Qiang2023, Gao2023, Sim2023, Sim2023_2, Zhang2024, Zhang2024-2, Potts2024, McGinley2024, McGinley2024, Huang2024}.

In particular, we establish how 2DCS can reveal quadrupolar excitations, using a spin-1 ferromagnetic Heisenberg model with single-ion anisotropy as an example case study. Using a combination of exact diagonalization (ED) and linear generalized spin-wave theory (GSWT), we show that the energy of the SIBS can be measured by the third-order non-linear susceptibility, $\chi^{(3;1)}_{xxxx}(\omega_t, \omega_\tau)$. Furthermore, we show that the other third-order non-linear susceptibility, $\chi^{(3;2)}(\omega_t, \omega_\tau)$, can provide a measure of the quadrupolar weight of an excitation, even in cases with finite hybridization between dipolar and quadrupolar excitations, information not discernible from linear response alone.

\textit{Model and its Excitations.---} As a paradigmatic example, we study the spin-1 ferromagnetic Heisenberg model with single-ion anisotropy on a regular $d$-dimensional lattice,
\begin{equation}
    H = - J \sum_{\avg{i,j}} \vec{S}_i \cdot \vec{S}_{j} - D \sum_i \left( S_i^z\right)^2,
    \label{eq:main_Hamiltonian}
\end{equation}
where $J>0$ is the ferromagnetic Heisenberg coupling, and $D>0$ the single ion-anisotropy~\cite{Silberglitt1970, Papanicolaou1987}. The model has a $U(1)\times\mathbb{Z}_2$ symmetry and the ground state, for $D/J>0$, is a fully polarized ferromagnet along the $z$-axis, $\ket{\Psi_0^\pm}=\ket{\pm1}^{\otimes N}$, spontaneously breaking the discrete $\mathbb{Z}_2$ symmetry. 

The $U(1)$ symmetry allows us to divide the Hilbert space into sectors with fixed $M^z=\sum_i S_i^z$. Since the application of $M^x \equiv \sum_i S^x_i$ can only change $M^z$ by $\pm 1$, focusing solely on the $M^z = N, N-1, N-2$ sectors is sufficient when considering zero-temperature 2DCS. Combined with translational symmetry, the dimension of the Hilbert space to be considered is thus $\mathcal{O}(N)$, enabling us to evaluate non-linear response using ED up to relatively large system sizes of $\mathcal{O}(100)$ sites (see Ref.~\cite{Watanabe2024} for details on the ED approach to 2DCS).

The energy spectrum at zero momentum, and its various excitations, calculated from ED for such a spin-1 ferromagnet is shown in Fig.~\ref{fig:chi1_Ddep}. The simplest elementary excitation is a single magnon, a dipolar $\abs{\Delta M^z}=1$ excitation created by applying $M^x$ to the fully polarized ground state. It has an energy $\omega_{\textrm{1m}}=D$ relative to the ground state. Above this, there is a continuum of two-magnon states, spanning an energy range $2D \leq \omega_\textrm{2m} \leq 2D+4zJ$, with $z$ the coordination number of the lattice. These are $\abs{\Delta M^z}=2$ composite excitations which consist of pairs of dipolar single magnon excitations. Schematically, for say the $\ket{\Psi_0^+}$ ground state, the 1-magnon and 2-magnon excitations can be understood as flipping $\ket{+1}_i\rightarrow\ket{0}_i$ on a single site and flipping $\ket{+1}_i\ket{+1}_j\rightarrow\ket{0}_i\ket{0}_j$ on two different sites, respectively.    

Crucially, for the $S=1$ model, there is an additional elementary excitation with an intrinsic multipolar character. This is the single-ion bound state (SIBS)~\cite{Silberglitt1970}, which, in the limit $D/J\rightarrow\infty$, can be understood as a full $\abs{\Delta M^z}=2$ spin flip on a single site, e.g.~$\ket{+1}_i \rightarrow \ket{-1}_i$. In this limit, the energy of the SIBS $\omega_\text{SIBS} \rightarrow 2zJ$. Note that such an on-site quadrupolar excitation cannot occur in a purely spin-$1/2$ system, and is furthermore completely absent in linear response, which is only sensitive to dipolar excitations. 
At {\it finite} $D/J$ it is important to realize that the nature of the SIBS is no longer as simple as the schematic $\ket{+1}_i \rightarrow \ket{-1}_i$ full spin flip presented in Fig.~\ref{fig:chi1_Ddep}(b) (such a simple excited state is not even an eigenstate of the Hamiltonian due to the XY exchange terms contained within the Heisenberg interactions). We can dissect its true character by expanding $(M^x)^2$ as
\begin{align}
 \notag  (M^x)^2 =& \frac{1}{2} \sum_i Q_i^{x^2-y^2} + \frac{1}{2}\sum_{i}(S_i^+ S_{i+1}^+ + S_i^- S_{i+1}^-) \\ &+ \frac{1}{4}\sum_{i\neq j,j+1}(S_i^+ S_j^+ + S_i^- S_j^-)  + \dots \,,
 \label{eq:Mx2}
\end{align}
where the first term represents a purely on-site quadrupolar excitation (unique to spin-1 systems), the second represents a 2-magnon bound state (two magnons bound on neighboring sites), the third a regular 2-magnon state, and the ellipses $\abs{\Delta M^z}=0$ terms which do not play a role for the SIBS at the moment. Taking $D/J=5$ in the 1-d chain as an example, the largest contribution for the SIBS  comes from the first term, $\sim 65\%$ of the total weight, while the second and third terms contribute $\sim 30\%$ and $\sim 5\%$ respectively (see the Supplementary Information for how the weights evolve as a function of $D/J$). On the other hand, the lowest energy state in the 2-magnon continuum is fully dominated by the third term, $\sim 99\%$, which creates two magnons on two different sites. Thus, though the SIBS and 2-magnon states are both $\abs{\Delta M^z}=2$ quadrupolar excitations, they have a fundamentally distinct underlying nature. The SIBS is a single, sharp elementary excitation while the 2-magnon states form a continuum, and the SIBS has primarily on-site quadrupolar character while the 2-magnon states are composed of pairs of single magnon excitations. At small enough $D/J$, the SIBS merges with the 2-magnon continuum and loses its distinct nature.


\begin{figure}
  \centering
  \includegraphics[width=1.0\columnwidth]{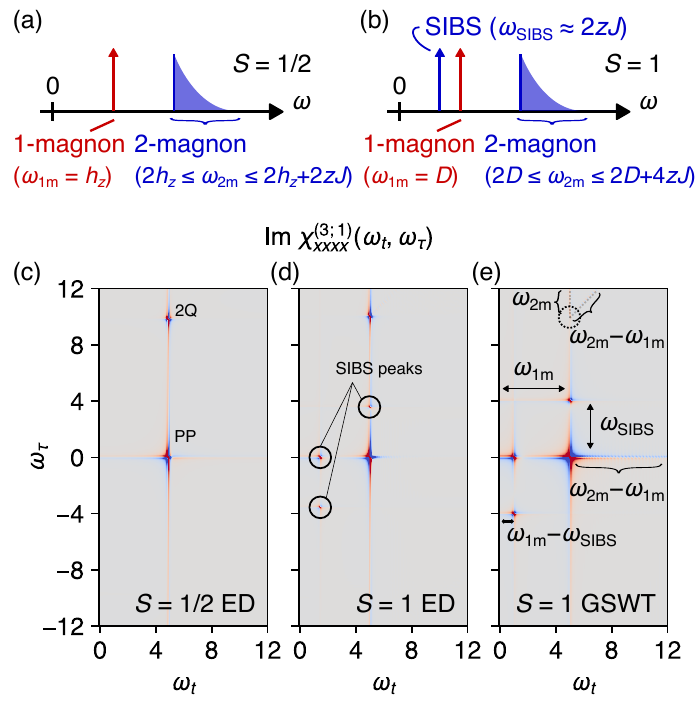}
  \caption{{\bf 2DCS signatures of quadrupolar excitations.}
  Schematic of the excitation energies in the large-$D$ limit for (a) the spin-1/2 FM XXZ model with longitudinal field and (b) the spin-1 FM Heisenberg model with single-ion anisotropy. 
  The non-linear third-order susceptibility, $\chi^{(3;1)}_{xxxx}$, calculated for (c) the spin-1/2 model with ED, and for the spin-1 model at $D/J =5.0$ with ED results shown in (d) and linear generalized spin wave theory results in (e). As a representative example, the 2DCS results are shown here for a 1D chain of length $L=100$. 
  }
  \label{fig:largeD}
\end{figure}

\textit{2DCS Quadrupolar Signatures.---} 
Having understood the structure of the excitation spectrum, we now turn to how 2DCS can be used to reveal the presence of the SIBS. We focus on the third-order diagonal susceptibility $\chi^{(3)}_{xxxx}(t_3,t_2,t_1)$ (the second-order diagonal susceptibility vanishes). In the two-pulse setup used for 2DCS, there are two distinct contributions, $\chi^{(3;1)}_{xxxx}(t,\tau,0)$ and $\chi^{(3;2)}_{xxxx}(t,0,\tau)$. For now, we focus on the Fourier transformed $\chi^{(3;1)}_{xxxx}(\omega_t,\omega_\tau)$, which corresponds to two field interactions at time $t^\prime=0$ and one at $t^\prime=\tau$.    

As a useful comparison, we first consider a spin-1/2 XXZ model with longitudinal field, with Hamiltonian $H = -J \sum_{\avg{i,j}}[\gamma(S_i^xS_{j}^x + S_i^yS_{j}^y)+ S_i^zS_{j}^z ] - h_z\sum_i S_i^z$, and $\gamma/J = 1.1, h_z/J = 5.0$. The ground state is also a fully polarized ferromagnet along the $z$-axis, with a 1-magnon excitation with energy $\omega_\text{1m} \approx h_z$ and a 2-magnon continuum for $\omega_\text{2m} \geq 2\omega_\text{1m}$ [Fig.~\ref{fig:largeD}(a)]. As shown in Fig.~\ref{fig:largeD}(c), in the 2DCS spectrum, there is a pump-probe (PP) peak at $(\omega_t, \omega_\tau) = (\omega_1, 0)$ due to the 1-magnon excitation~\cite{Lu2017}. In addition, the lower edge of the 2-magnon continuum is visible as a 2Q peak at $(\omega_1, 2\omega_1)$~\cite{Lu2017}. 

Now we consider our $S=1$ model with $D/J\gg 1$. In this limit, as schematically shown in Fig.~\ref{fig:largeD}(b), the SIBS is well separated from both the 1-magnon excitation and the 2-magnon continuum. In the 2DCS spectrum, we again observe the same PP and 2Q peaks from the 1-magnon excitation and 2-magnon continuum respectively, just as in the spin-$1/2$ model. However, $\chi^{(3;1)}_{xxxx}$ now also contains {\it three additional peaks}, roughly located at $(\omega_\text{1m}, \omega_{\text{SIBS}})$, $(\omega_\text{1m} - \omega_{\text{SIBS}}, \omega_{\text{SIBS}})$, and $(\omega_\text{1m} - \omega_{\text{SIBS}}, 0)$. 
These additional peaks reveal quadrupolar excitation processes as can be intuitively understood, for instance, for the first of these peaks as follows:
At time $t^\prime=0$, the two field interactions generate a SIBS excitation with energy $\omega_\text{SIBS}$ via $M^x(0)M^x(0)$. The system then evolves for a time $\tau$, generating a peak at $\omega_\tau=\omega_\text{SIBS}$. At time $t^\prime=\tau$, the single field interaction demotes the SIBS to a single magnon excitation via $M^x(\tau)$, which, after the system evolves for a time $t$, generates a peak at $\omega_t=\omega_\text{1m}$. Finally, at time $t^\prime=t+\tau$, the system returns to the ground state via $M^x(t+\tau)$. 

A more intuitive understanding of the 2DCS spectra can be obtained by comparing the numerical ED calculations with the linear generalized spin wave theory (GSWT) approach~\cite{Bai2021}, which treats magnon and quadrupolar excitations on equal footing. In GSWT, the Hamiltonian can be re-expressed using an SU(3) Schwinger boson $(\beta_{i,+1}, \beta_{i,0}, \beta_{i,-1})$ representation. In the FM case, the exact ground state, which we take here to be the fully polarized state $\ket{\Psi_0^+}=\ket{+1}^{\otimes N}$, can be considered as a condensation of the $\beta_{i,+1}$ bosons. The remaining two flavors of bosons, $\beta_{i,0}$ and $\beta_{i,-1}$ correspond to $\ket{0}_i$ and $\ket{-1}_i$ on-site excitations, which can then be used to represent the $\beta_{i,+1}$ operator as $\beta^\dagger_{i,+1} = \beta_{i,+1} = (1 - \beta^\dagger_{i,0}\beta_{i,0} - \beta^\dagger_{i,-1}\beta_{i,-1})^{1/2}$. Expanding the square root to quadratic order in $\beta_{i,0}, \beta_{i,-1}$, and inserting these into Eq.~(\ref{eq:main_Hamiltonian}), we obtain the linear GSWT Hamiltonian in momentum space, for a lattice with a single-site unit cell, as
\begin{equation}
  H_{\text{GSWT}} = \sum_{k} \omega_k \beta^\dagger_{k,0} \beta^{\phantom\dagger}_{k,0} + 4J \sum_k \beta^\dagger_{k,1}\beta^{\phantom\dagger}_{k,1},
\end{equation}
where $\omega_k = zJ + D - J \gamma_\vec{k}$, with $\gamma_\vec{k} = \sum_\delta e^{i\vec{k}\cdot\vec{e}_\delta}$ and $\vec{e}_\delta$ the $z$ nearest-neighbor vectors. At $k=0$, the energies of the 1-magnon and 2-magnon states are identical to the numerical ED results. On the other hand, the energy of the SIBS $\omega_{\text{SIBS}}$ is given as exactly $4J$, which agrees with the ED only in the limit $D/J\gg 1$. The inclusion of higher-order terms in the expansion of the square root is thus necessary to obtain the correct energy of the SIBS at smaller $D/J$. Indeed, the XY exchange part of the Hamiltonian produces a quartic term proportional to $M^0$ which, following a mean-field decomposition, will result in a correction to the SIBS dispersion. Using the GSWT approach, we can calculate the third-order susceptibility $\chi^{(3)}_{xxxx}(t,\tau,0)$ in the time domain using
\begin{align}
\notag \chi^{(3)}_{xxxx}(t,\tau,0) =
 -\frac{1}{N}\sum_{PQR} &A_{PQR}\left[
  2\sin\left(  \Delta E_{PR} \tau + \Delta E_{PQ} t \right)\right. \\
\notag  &  + \sin\left( - E_{Q} \tau + \Delta E_{PQ} t \right) \\
 & \left. +   \sin\left(  E_{Q} \tau + E_R t \right)  \right] \,,
  \label{eq:chi31}
\end{align}
where we have defined $\Delta E_{nm} = E_n - E_m$ (for simplicity we have set $E_0=0$), and $A_{PQR} = \braket{0|M^x|P}\braket{P|M^x|Q}\braket{Q|M^x|R}\braket{R|M^x|0}$ is the transition amplitude. The result, shown in Fig.~\ref{fig:largeD}(d), demonstrates that, apart from the 2Q peak, the GSWT calculations successfully reproduce all of the qualitative features of the ED results. Indeed, all three peaks unique to the spin-1 model are confirmed to originate from processes in which the intermediate state $\ket{Q}$ is the SIBS $\beta_{k=0, -1}^\dagger\ket{\text{vac.}}$. Not only does the GSWT approach provide insights into the origins of the various peaks, much more importantly it can also be straightforwardly applied to other magnetically ordered states.  

The one discrepancy, the absence of the 2Q peak, can be easily explained. In linear GSWT, the lowest energy state in the 2-magnon continuum, a pair of zone-center magnons, has exactly twice the energy of the 1-magnon state. As a result, each contribution to the 2Q peak, i.e.~the second and third terms in Eq.~(\ref{eq:chi31}), where $\ket{Q}$ is the 2-magnon state and $\ket{P}, \ket{Q}$ are 1-magnon states, cancel each other out. This cancellation leads to the absence of a pronounced 2Q peak. Conversely, the ED calculations reveal a finite shift in the lower edge of the 2-magnon continuum away from precisely $2\omega_\text{1m}$, which accounts for the 2Q peak observed in the 2DCS.

Finally, we should also note that a system size-independent PP peak is a result of an imperfect cancellation of two extensive contributions, i.e.~the processes where $\ket{Q} = \ket{0}$ and $\ket{Q} = (1/\sqrt{2})(\beta_{k=0}^\dagger)^2\ket{\text{vac.}}$ (pair of zone-center magnons). In ED, the shift in the 2-magnon energy results in a slight shift of the two contributions on top of the imperfect cancellation, leading to a diverging signal in the time domain at long times~\cite{Watanabe2024}. Since the SIBS signal is not divergent in the time domain, the choice of time window used for the Fourier transform affects the relative strength of the peaks in the frequency domain.

\begin{figure}
  \centering
  \includegraphics[width=1.0\columnwidth]{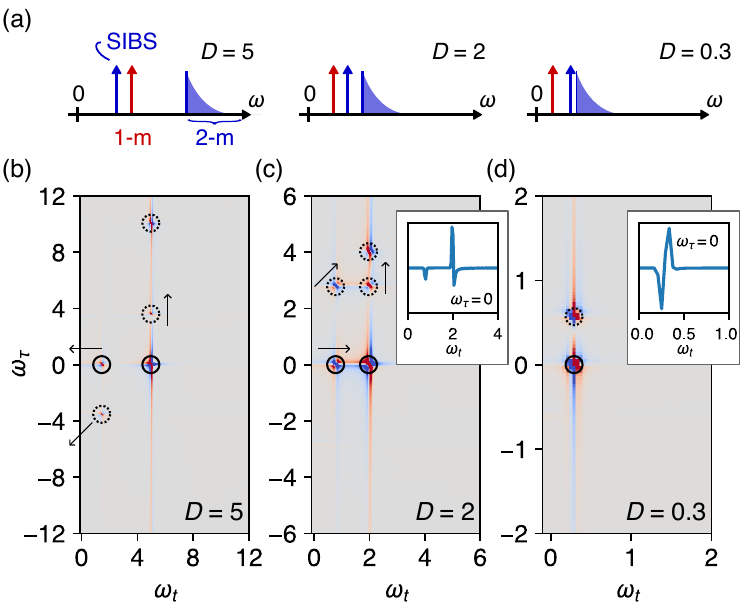}
  \caption{{\bf Evolution of excitations}. 
  	Dependence of $\chi^{(3;1)}_{xxxx}(\omega_t, \omega_\tau)$ for $D/J = 5.0, 2.0$ and $0.3$ obtained by ED for an $L = 100$ chain. 
  As schematically shown above each plot, the relative position of the 1-magnon, SIBS, and 2-magnon continuum changes as $D/J$ decreases. Arrows near the three SIBS peaks indicate the directions in which these peaks shift towards the reference points at $(\omega_1, 0)$ and $(\omega_1, 2\omega_1)$ as $D/J$ decreases. All of the peaks highlighted with solid circles eventually merge into one, likewise for those highlighted with dotted circles. Insets show the line cut along the horizontal axis $\omega_\tau = 0$.}
  \label{fig:chi31_Ddep}
\end{figure}

\textit{Evolution of Excitations.---} 
So far, we have focused on the limit $D/J\gg 1$. Let us now consider what happens to the 2DCS spectrum as we lower the value of $D/J$. As $D/J$ decreases, both $\omega_\text{1m}$ and $\omega_{\text{SIBS}}$ decrease, with their relative difference $\omega_\text{1m}-\omega_{\text{SIBS}}$ changing sign at, e.g.~for the 1D chain, $D \approx 3$ (see the Supplementary Information for 2D and 3D cases).
Consequently, the 2DCS peaks move within the 2D frequency plane, as indicated by the arrows in Fig.~\ref{fig:chi31_Ddep}(b) and (c). Further decreasing $D/J$ brings $\omega_{\text{SIBS}}$ close to the lower edge of the 2-magnon continuum, and mixing between the states results in the SIBS acquiring dominant 2-magnon character. In this limit, the 2DCS spectrum now resembles that of the spin-1/2 model, in which only PP and 2Q peaks are visible [Fig.~\ref{fig:chi31_Ddep}(d)]. However, as the PP peak consists of contributions from both the SIBS and the 1-magnon, an $\omega_\tau = 0$ line cut reveals a sign change at approximately $\omega_t \approx \omega_\text{1m} \approx \omega_{\text{SIBS}} - \omega_\text{1m}$ [insets of Fig.~\ref{fig:chi31_Ddep}]. 

Note that the small $D/J$ regime is of direct relevance to the material NiNb$_2$O$_6$, considered to be an experimental realization of a spin-1 chain with easy-axis single-ion anisotropy~\cite{Chauhan2020}. There, the ratio $D/J$ is estimated to be around $0.3$~\cite{Chauhan2020}, which would put the material into the regime described above in which the SIBS has a strong 2-magnon character and the unique spin-1 peaks cannot be easily discerned [Fig.~\ref{fig:chi31_Ddep}(d)]. 

\begin{figure}
  \centering
  \includegraphics[width=1.0\columnwidth]{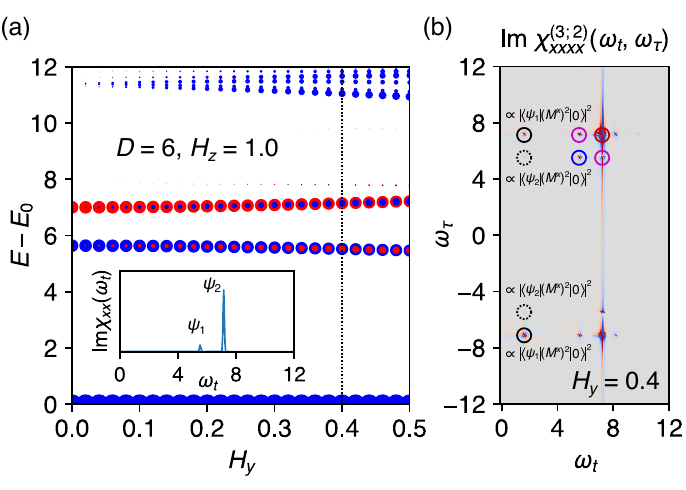}
  \caption{{\bf Extracting quadrupolar weight of hybridized magnon and SIBS modes}. (a) The evolution of the energy spectrum as a function of $H_y$ obtained by ED with $L=12$. The size of circles are proportional to $|\braket{n|M^x|0}|$ (red) and $|\braket{n|(M^x)^2|0}|$ (blue). Inset shows the linear response at $H_y = 0.4$, where two modes $\ket{\psi_1}$ and $\ket{\psi_2}$ are visible due to the $H_y$ induced hybridization. (b) Third order response $\chi^{(3;2)}_{xxxx}(\omega_t, \omega_\tau)$ at $H_y = 0.4$. In addition to diagonal peaks of $\ket{\psi_1}$ (blue), $\ket{\psi_2}$ (red), and off-diagonal peaks (pink, suggesting the hybridized nature of the two excitations), four peaks at $(\omega_\tau, \omega_\tau) = (E_2 - E_1, \pm E_1)$ and $(E_2 - E_1, \pm E_2)$ are observed. These peaks measure the quadrupolar weight of the modes.}
  \label{fig:hybridized}
\end{figure}

\textit{Quadrupolar Weight.---}  
In low symmetry models, in which the SIBS can hybridize with a single magnon, the resulting hybridized mode can already be observed with conventional linear response probes such as neutron scattering~\cite{Legros2021, Bai2021, Bai2023}. As an example, consider the effect of a tilted magnetic field, $-h_z\sum_i S_i^z - h_y\sum_i S_i^y$, on our FM spin-1 Heisenberg model. 
The transverse field $h_y$ breaks the $U(1)$ symmetry of the model and thereby hybridizes the SIBS and the 1-magnon excitation, generating two modes, $\ket{\psi_1}$ and $\ket{\psi_2}$, which are both visible within linear response (the longitudinal field $h_z$ is simply added to prevent, for the 1-d chain, the creation of free propagating domain walls on either side of the SIBS). The intensity of the linear response peaks, given by the matrix element $\abs{\bra{\psi_n} M^x \ket{0}}^2$ and shown in Fig.~\ref{fig:hybridized}(a), is a measure of the dipolar weight of the excitations. We fix $D/J=5$ such that, in the limit $h_y\rightarrow 0$, the modes $\ket{\psi_1}$ and $\ket{\psi_2}$ smoothly connect to the SIBS and 1-magnon excitation respectively.   

We can now use 2DCS to obtain additional information on the nature of the hybridized excitations, not accessible within linear response. To do so, we utilize the other third-order susceptibility $\chi^{(3;2)}_{xxxx}(\omega_t, \omega_\tau) \equiv \text{FT}[\chi^{(3)}_{xxxx}(t, 0, \tau)]$. As shown in Fig.~\ref{fig:hybridized}(b), the diagonal nonrephasing peaks of $\ket{\psi_1}$ and $\ket{\psi_2}$ are clearly visible, and the appearance of cross peaks points to the hybridized nature of the two excitations~\cite{Hamm_Zanni_2011}. In addition, four peaks at $(\omega_\tau, \omega_\tau) = (\omega_2 - \omega_1, \pm \omega_1)$ and $(\omega_2 - \omega_1, \pm \omega_2)$ are observed. One can show that the intensities of these additional peaks are given by
\begin{equation}
  \begin{split}
    I_{(\omega_2 - \omega_1, \pm \omega_1)} \propto m_{01}^x m_{12}^x m_{21}^x m_{10}^x \approx \left|\braket{\psi_2|(M^x)^2|0}\right|^2,\\
    I_{(\omega_2 - \omega_1, \pm \omega_2)} \propto m_{02}^x m_{21}^x m_{12}^x m_{20}^x \approx \left|\braket{\psi_1|(M^x)^2|0}\right|^2,
  \end{split}
\end{equation}
where $m_{nm}^x = \braket{n|M^x|m}$, and the last approximation is justified since $m_{nm}^x$ for $n > 2$ is negligible and $m_{nn}^x = 0$. We numerically confirm that the deviation between the intensity $I_{(E_2 - E_1, \pm E_1)}$ and $\left|\braket{\psi_2|(M^x)^2|0}\right|^2$ is less than $10\%$ for $h_x < 0.5$ (and similar for the other peak), with the deviation largely due to contributions from higher excited states. Thus, the intensity of these 2DCS peaks can provide a measure of the quadrupolar weight of the excitations, $\left|\braket{\psi_n|(M^x)^2|0}\right|^2$, analogous to how the intensity of the linear response peaks provides a measure of their dipolar weight, $\left|\braket{\psi_n|M^x|0}\right|^2$, allowing for a more sophisticated characterization of the quasiparticle spectrum. 

We can again contrast this with the spin-$1/2$ XXZ model from before, with an additional staggered tilted field that breaks U(1) symmetry and produces two distinct dipolar magnon excitations (see Supplementary for more details). In that case, hybridization between the magnon modes occurs due to the exchange interaction but neither mode possesses quadrupolar weight, meaning that the additional peaks are absent. 

Note that in the case of more than two modes, we expect to observe sets of peaks at $(|E_n - E_m|, \pm E_n)$. 
The intensity of individual peaks is then proportional to $|m_{nm}^x|^2 | m_{0n}^x|^2$, providing information on the dipolar transition amplitudes between the excited states $n$ and $m$, as well as the usual dipolar transition amplitude from the ground state.

\textit{Outlook.--}  
For various magnetic materials quadrupolar excitations have remained a theoretically expected, but typically experimentally hidden feature, which 2DCS should now be able to clearly uncover. 
One material of much current interest is the easy-axis triangular magnet FeI$_2$ -- another effective spin-1 magnet where the observation of a hybridized SIBS in neutron scattering experiment has been discussed in great detail~\cite{Bai2021, Bai2023}. Our work suggests that non-linear spectroscopy could be used to directly probe the putative quadrupolar nature of these excitations -- in particular, since for FeI$_2$ the SIBS 
is better separated from the 2-magnon continuum than for the NiNb$_2$O$_6$ chain compound mentioned previously. 
Along similar lines, one might reinspect other Ni$^{2+}$ compounds \cite{Koch2003} which may also show well-separated SIBS.

Outside of higher-spin magnets, another interesting class of materials expected to exhibit quadrupolar excitations are systems with non-Kramers doublets~\cite{Lee2012, Liu2018}
whose local moments carry quadrupolar character. This includes the rare-earth intermetallics  Pr(Ti,V,Ir)$_2$(Al,Zn)$_{20}$ which exhibit quadrupolar-octupolar non-Kramers doublets~\cite{Sakai2011,Freyer2018} or $5d^2$ vacancy-ordered halide double perovskites~\cite{Pradhan2024}.
As these systems lack any dipolar character, their excitations remain invisible in many conventional probes, but non-linear 2DCS probes could be used to reveal their quadrupolar excitations. 
Indeed, the spin-1 FM studied here can be alternatively understood exactly within this framework. 
With $D\gg J$, the low-energy Hilbert space consists of a single pseudospin doublet, with $\ket{\uparrow} = \ket{+1}$ and $\ket{\downarrow} = \ket{-1}$, while the state $\ket{0}$ is separated by an energy cost proportional to $D$. One can then define associated pseudospin operators: (i) $\tau_i^z = S_i^z/2$, which transforms as a dipolar, time-reversal odd operator, and (ii) $\tau_i^\pm = S_i^\pm S_i^\pm$, which transform as quadrupolar, time-reversal even operators. The effective Hamiltonian, at lowest order, is simply a classical FM Ising model, $H_{\textrm{eff}}=-2J\sum_i \tau_i^z \tau_{i+1}^z$. The excitations are localized spin-flips created by $\tau_i^\pm$, with energy cost $4J$. But since $\tau_i^\pm$ carries purely quadrupolar character, these excitations remain invisible in linear response. However, they can be observed with 2DCS, such as in $\chi_{xxxx}^{(3;1)}$ as discussed here, albeit only through intermediate states outside the low-energy doublet Hilbert space (which means they only appear in a region of the 2D frequency space at distances on the order of $D$ from the origin). 

In general, it should be clear that non-linear response has enormous potential in probing multipolar excitations, both elementary mutipolar excitations in $S > 1/2$ systems 
with easy-axis anisotropy or field-induced polarized phases~\cite{Zvyagin2008, Akaki2017}, and in non-Kramers doublets with multipolar character, as well as composite multipolar excitations such as exchange bound states, feasible even in spin-1/2 systems 
\cite{Coldea2010, Nishida2013, Morris2014, Grenier2015, Wang2018, Dally2020, Sim2023}. Though we focused here on a relatively simple spin-1 FM as a proof-of-concept example in this work, the linear GSWT approach can be straightforwardly extended to other ordered phases. In particular, in cases where a single magnetic field pulse only creates single quasiparticle excitations, taking up to the two-particle subspace into account should suffice to compute the 2DCS spectrum, with non-linear corrections able to be added either through a mean-field approach or direct numerical diagonalization~\cite{Bai2023}.  
\\

\textit{Data availability.---}
The numerical data shown in the figures is available on Zenodo~\cite{zenodo_repository}.  \\

\textit{Acknowledgments.---} 
We thank A.~Paramekanti for illuminating discussions. 
We acknowledge partial funding from the DFG within Project-ID 277146847, SFB 1238 (projects C02, C03). 
This work has benefitted from multiple exchanges during the 2023 KITP workshop ``A New Spin on Quantum Magnets", 
supported in part by grant NSF PHY-1748958 to the Kavli Institute for Theoretical Physics (KITP).
The numerical simulations were performed on the JUWELS cluster at the Forschungszentrum Juelich 
and the Noctua2 cluster at PC2 in Paderborn.


\bibliography{2DCS}

\begin{thebibliography}{46}%
\makeatletter
\providecommand \@ifxundefined [1]{%
 \@ifx{#1\undefined}
}%
\providecommand \@ifnum [1]{%
 \ifnum #1\expandafter \@firstoftwo
 \else \expandafter \@secondoftwo
 \fi
}%
\providecommand \@ifx [1]{%
 \ifx #1\expandafter \@firstoftwo
 \else \expandafter \@secondoftwo
 \fi
}%
\providecommand \natexlab [1]{#1}%
\providecommand \enquote  [1]{``#1''}%
\providecommand \bibnamefont  [1]{#1}%
\providecommand \bibfnamefont [1]{#1}%
\providecommand \citenamefont [1]{#1}%
\providecommand \href@noop [0]{\@secondoftwo}%
\providecommand \href [0]{\begingroup \@sanitize@url \@href}%
\providecommand \@href[1]{\@@startlink{#1}\@@href}%
\providecommand \@@href[1]{\endgroup#1\@@endlink}%
\providecommand \@sanitize@url [0]{\catcode `\\12\catcode `\$12\catcode `\&12\catcode `\#12\catcode `\^12\catcode `\_12\catcode `\%12\relax}%
\providecommand \@@startlink[1]{}%
\providecommand \@@endlink[0]{}%
\providecommand \url  [0]{\begingroup\@sanitize@url \@url }%
\providecommand \@url [1]{\endgroup\@href {#1}{\urlprefix }}%
\providecommand \urlprefix  [0]{URL }%
\providecommand \Eprint [0]{\href }%
\providecommand \doibase [0]{https://doi.org/}%
\providecommand \selectlanguage [0]{\@gobble}%
\providecommand \bibinfo  [0]{\@secondoftwo}%
\providecommand \bibfield  [0]{\@secondoftwo}%
\providecommand \translation [1]{[#1]}%
\providecommand \BibitemOpen [0]{}%
\providecommand \bibitemStop [0]{}%
\providecommand \bibitemNoStop [0]{.\EOS\space}%
\providecommand \EOS [0]{\spacefactor3000\relax}%
\providecommand \BibitemShut  [1]{\csname bibitem#1\endcsname}%
\let\auto@bib@innerbib\@empty
\bibitem [{\citenamefont {Penc}\ \emph {et~al.}(2012)\citenamefont {Penc}, \citenamefont {Romh\'anyi}, \citenamefont {R\~o\ om}, \citenamefont {Nagel}, \citenamefont {Antal}, \citenamefont {Feh\'er}, \citenamefont {J\'anossy}, \citenamefont {Engelkamp}, \citenamefont {Murakawa}, \citenamefont {Tokura}, \citenamefont {Szaller}, \citenamefont {Bord\'acs},\ and\ \citenamefont {K\'ezsm\'arki}}]{Penc2012}%
  \BibitemOpen
  \bibfield  {author} {\bibinfo {author} {\bibfnamefont {K.}~\bibnamefont {Penc}}, \bibinfo {author} {\bibfnamefont {J.}~\bibnamefont {Romh\'anyi}}, \bibinfo {author} {\bibfnamefont {T.}~\bibnamefont {R\~o\ om}}, \bibinfo {author} {\bibfnamefont {U.}~\bibnamefont {Nagel}}, \bibinfo {author} {\bibfnamefont {A.}~\bibnamefont {Antal}}, \bibinfo {author} {\bibfnamefont {T.}~\bibnamefont {Feh\'er}}, \bibinfo {author} {\bibfnamefont {A.}~\bibnamefont {J\'anossy}}, \bibinfo {author} {\bibfnamefont {H.}~\bibnamefont {Engelkamp}}, \bibinfo {author} {\bibfnamefont {H.}~\bibnamefont {Murakawa}}, \bibinfo {author} {\bibfnamefont {Y.}~\bibnamefont {Tokura}}, \bibinfo {author} {\bibfnamefont {D.}~\bibnamefont {Szaller}}, \bibinfo {author} {\bibfnamefont {S.}~\bibnamefont {Bord\'acs}},\ and\ \bibinfo {author} {\bibfnamefont {I.}~\bibnamefont {K\'ezsm\'arki}},\ }\bibfield  {title} {\bibinfo {title} {{Spin-Stretching Modes in Anisotropic Magnets: Spin-Wave Excitations in the Multiferroic Ba$_2$CoGe$_2$O$_7$}},\ }\href {https://doi.org/10.1103/PhysRevLett.108.257203} {\bibfield  {journal} {\bibinfo  {journal} {Phys. Rev. Lett.}\ }\textbf {\bibinfo {volume} {108}},\ \bibinfo {pages} {257203} (\bibinfo {year} {2012})}\BibitemShut {NoStop}%
\bibitem [{\citenamefont {Bai}\ \emph {et~al.}(2021)\citenamefont {Bai}, \citenamefont {Zhang}, \citenamefont {Dun}, \citenamefont {Zhang}, \citenamefont {Huang}, \citenamefont {Zhou}, \citenamefont {Stone}, \citenamefont {Kolesnikov}, \citenamefont {Ye}, \citenamefont {Batista},\ and\ \citenamefont {Mourigal}}]{Bai2021}%
  \BibitemOpen
  \bibfield  {author} {\bibinfo {author} {\bibfnamefont {X.}~\bibnamefont {Bai}}, \bibinfo {author} {\bibfnamefont {S.-S.}\ \bibnamefont {Zhang}}, \bibinfo {author} {\bibfnamefont {Z.}~\bibnamefont {Dun}}, \bibinfo {author} {\bibfnamefont {H.}~\bibnamefont {Zhang}}, \bibinfo {author} {\bibfnamefont {Q.}~\bibnamefont {Huang}}, \bibinfo {author} {\bibfnamefont {H.}~\bibnamefont {Zhou}}, \bibinfo {author} {\bibfnamefont {M.~B.}\ \bibnamefont {Stone}}, \bibinfo {author} {\bibfnamefont {A.~I.}\ \bibnamefont {Kolesnikov}}, \bibinfo {author} {\bibfnamefont {F.}~\bibnamefont {Ye}}, \bibinfo {author} {\bibfnamefont {C.~D.}\ \bibnamefont {Batista}},\ and\ \bibinfo {author} {\bibfnamefont {M.}~\bibnamefont {Mourigal}},\ }\bibfield  {title} {\bibinfo {title} {{Hybridized quadrupolar excitations in the spin-anisotropic frustrated magnet FeI$_2$}},\ }\href {https://doi.org/10.1038/s41567-020-01110-1} {\bibfield  {journal} {\bibinfo  {journal} {Nature Physics}\ }\textbf {\bibinfo {volume} {17}},\ \bibinfo {pages} {467} (\bibinfo {year} {2021})}\BibitemShut {NoStop}%
\bibitem [{\citenamefont {Remund}\ \emph {et~al.}(2022)\citenamefont {Remund}, \citenamefont {Pohle}, \citenamefont {Akagi}, \citenamefont {Romh\'anyi},\ and\ \citenamefont {Shannon}}]{Remund2022}%
  \BibitemOpen
  \bibfield  {author} {\bibinfo {author} {\bibfnamefont {K.}~\bibnamefont {Remund}}, \bibinfo {author} {\bibfnamefont {R.}~\bibnamefont {Pohle}}, \bibinfo {author} {\bibfnamefont {Y.}~\bibnamefont {Akagi}}, \bibinfo {author} {\bibfnamefont {J.}~\bibnamefont {Romh\'anyi}},\ and\ \bibinfo {author} {\bibfnamefont {N.}~\bibnamefont {Shannon}},\ }\bibfield  {title} {\bibinfo {title} {Semi-classical simulation of spin-1 magnets},\ }\href {https://doi.org/10.1103/PhysRevResearch.4.033106} {\bibfield  {journal} {\bibinfo  {journal} {Phys. Rev. Res.}\ }\textbf {\bibinfo {volume} {4}},\ \bibinfo {pages} {033106} (\bibinfo {year} {2022})}\BibitemShut {NoStop}%
\bibitem [{\citenamefont {Silberglitt}\ and\ \citenamefont {Torrance}(1970)}]{Silberglitt1970}%
  \BibitemOpen
  \bibfield  {author} {\bibinfo {author} {\bibfnamefont {R.}~\bibnamefont {Silberglitt}}\ and\ \bibinfo {author} {\bibfnamefont {J.~B.}\ \bibnamefont {Torrance}},\ }\bibfield  {title} {\bibinfo {title} {{Effect of Single-Ion Anisotropy on Two-Spin-Wave Bound State in a Heisenberg Ferromagnet}},\ }\href {https://doi.org/10.1103/PhysRevB.2.772} {\bibfield  {journal} {\bibinfo  {journal} {Phys. Rev. B}\ }\textbf {\bibinfo {volume} {2}},\ \bibinfo {pages} {772} (\bibinfo {year} {1970})}\BibitemShut {NoStop}%
\bibitem [{\citenamefont {Zhitomirsky}\ and\ \citenamefont {Chernyshev}(2013)}]{Zhitomirsky2013}%
  \BibitemOpen
  \bibfield  {author} {\bibinfo {author} {\bibfnamefont {M.~E.}\ \bibnamefont {Zhitomirsky}}\ and\ \bibinfo {author} {\bibfnamefont {A.~L.}\ \bibnamefont {Chernyshev}},\ }\bibfield  {title} {\bibinfo {title} {{Colloquium: Spontaneous magnon decays}},\ }\href {https://doi.org/10.1103/RevModPhys.85.219} {\bibfield  {journal} {\bibinfo  {journal} {Rev. Mod. Phys.}\ }\textbf {\bibinfo {volume} {85}},\ \bibinfo {pages} {219} (\bibinfo {year} {2013})}\BibitemShut {NoStop}%
\bibitem [{\citenamefont {Legros}\ \emph {et~al.}(2021)\citenamefont {Legros}, \citenamefont {Zhang}, \citenamefont {Bai}, \citenamefont {Zhang}, \citenamefont {Dun}, \citenamefont {Phelan}, \citenamefont {Batista}, \citenamefont {Mourigal},\ and\ \citenamefont {Armitage}}]{Legros2021}%
  \BibitemOpen
  \bibfield  {author} {\bibinfo {author} {\bibfnamefont {A.}~\bibnamefont {Legros}}, \bibinfo {author} {\bibfnamefont {S.-S.}\ \bibnamefont {Zhang}}, \bibinfo {author} {\bibfnamefont {X.}~\bibnamefont {Bai}}, \bibinfo {author} {\bibfnamefont {H.}~\bibnamefont {Zhang}}, \bibinfo {author} {\bibfnamefont {Z.}~\bibnamefont {Dun}}, \bibinfo {author} {\bibfnamefont {W.~A.}\ \bibnamefont {Phelan}}, \bibinfo {author} {\bibfnamefont {C.~D.}\ \bibnamefont {Batista}}, \bibinfo {author} {\bibfnamefont {M.}~\bibnamefont {Mourigal}},\ and\ \bibinfo {author} {\bibfnamefont {N.~P.}\ \bibnamefont {Armitage}},\ }\bibfield  {title} {\bibinfo {title} {{Observation of 4- and 6-Magnon Bound States in the Spin-Anisotropic Frustrated Antiferromagnet FeI$_{2}$}},\ }\href {https://doi.org/10.1103/PhysRevLett.127.267201} {\bibfield  {journal} {\bibinfo  {journal} {Phys. Rev. Lett.}\ }\textbf {\bibinfo {volume} {127}},\ \bibinfo {pages} {267201} (\bibinfo {year} {2021})}\BibitemShut {NoStop}%
\bibitem [{\citenamefont {Bai}\ \emph {et~al.}(2023)\citenamefont {Bai}, \citenamefont {Zhang}, \citenamefont {Zhang}, \citenamefont {Dun}, \citenamefont {Phelan}, \citenamefont {Garlea}, \citenamefont {Mourigal},\ and\ \citenamefont {Batista}}]{Bai2023}%
  \BibitemOpen
  \bibfield  {author} {\bibinfo {author} {\bibfnamefont {X.}~\bibnamefont {Bai}}, \bibinfo {author} {\bibfnamefont {S.-S.}\ \bibnamefont {Zhang}}, \bibinfo {author} {\bibfnamefont {H.}~\bibnamefont {Zhang}}, \bibinfo {author} {\bibfnamefont {Z.}~\bibnamefont {Dun}}, \bibinfo {author} {\bibfnamefont {W.~A.}\ \bibnamefont {Phelan}}, \bibinfo {author} {\bibfnamefont {V.~O.}\ \bibnamefont {Garlea}}, \bibinfo {author} {\bibfnamefont {M.}~\bibnamefont {Mourigal}},\ and\ \bibinfo {author} {\bibfnamefont {C.~D.}\ \bibnamefont {Batista}},\ }\bibfield  {title} {\bibinfo {title} {Instabilities of heavy magnons in an anisotropic magnet},\ }\href {https://doi.org/10.1038/s41467-023-39940-1} {\bibfield  {journal} {\bibinfo  {journal} {Nature Communications}\ }\textbf {\bibinfo {volume} {14}},\ \bibinfo {pages} {4199} (\bibinfo {year} {2023})}\BibitemShut {NoStop}%
\bibitem [{\citenamefont {Nag}\ \emph {et~al.}(2022)\citenamefont {Nag}, \citenamefont {Nocera}, \citenamefont {Agrestini}, \citenamefont {Garcia-Fernandez}, \citenamefont {Walters}, \citenamefont {Cheong}, \citenamefont {Johnston},\ and\ \citenamefont {Zhou}}]{Nag2022}%
  \BibitemOpen
  \bibfield  {author} {\bibinfo {author} {\bibfnamefont {A.}~\bibnamefont {Nag}}, \bibinfo {author} {\bibfnamefont {A.}~\bibnamefont {Nocera}}, \bibinfo {author} {\bibfnamefont {S.}~\bibnamefont {Agrestini}}, \bibinfo {author} {\bibfnamefont {M.}~\bibnamefont {Garcia-Fernandez}}, \bibinfo {author} {\bibfnamefont {A.~C.}\ \bibnamefont {Walters}}, \bibinfo {author} {\bibfnamefont {S.-W.}\ \bibnamefont {Cheong}}, \bibinfo {author} {\bibfnamefont {S.}~\bibnamefont {Johnston}},\ and\ \bibinfo {author} {\bibfnamefont {K.-J.}\ \bibnamefont {Zhou}},\ }\bibfield  {title} {\bibinfo {title} {Quadrupolar magnetic excitations in an isotropic spin-1 antiferromagnet},\ }\href {https://doi.org/10.1038/s41467-022-30065-5} {\bibfield  {journal} {\bibinfo  {journal} {Nature Communications}\ }\textbf {\bibinfo {volume} {13}},\ \bibinfo {pages} {2327} (\bibinfo {year} {2022})}\BibitemShut {NoStop}%
\bibitem [{\citenamefont {Lu}\ \emph {et~al.}(2017)\citenamefont {Lu}, \citenamefont {Li}, \citenamefont {Hwang}, \citenamefont {Ofori-Okai}, \citenamefont {Kurihara}, \citenamefont {Suemoto},\ and\ \citenamefont {Nelson}}]{Lu2017}%
  \BibitemOpen
  \bibfield  {author} {\bibinfo {author} {\bibfnamefont {J.}~\bibnamefont {Lu}}, \bibinfo {author} {\bibfnamefont {X.}~\bibnamefont {Li}}, \bibinfo {author} {\bibfnamefont {H.~Y.}\ \bibnamefont {Hwang}}, \bibinfo {author} {\bibfnamefont {B.~K.}\ \bibnamefont {Ofori-Okai}}, \bibinfo {author} {\bibfnamefont {T.}~\bibnamefont {Kurihara}}, \bibinfo {author} {\bibfnamefont {T.}~\bibnamefont {Suemoto}},\ and\ \bibinfo {author} {\bibfnamefont {K.~A.}\ \bibnamefont {Nelson}},\ }\bibfield  {title} {\bibinfo {title} {{Coherent Two-Dimensional Terahertz Magnetic Resonance Spectroscopy of Collective Spin Waves}},\ }\href {https://doi.org/10.1103/PhysRevLett.118.207204} {\bibfield  {journal} {\bibinfo  {journal} {Phys. Rev. Lett.}\ }\textbf {\bibinfo {volume} {118}},\ \bibinfo {pages} {207204} (\bibinfo {year} {2017})}\BibitemShut {NoStop}%
\bibitem [{\citenamefont {Wan}\ and\ \citenamefont {Armitage}(2019)}]{Wan2019}%
  \BibitemOpen
  \bibfield  {author} {\bibinfo {author} {\bibfnamefont {Y.}~\bibnamefont {Wan}}\ and\ \bibinfo {author} {\bibfnamefont {N.~P.}\ \bibnamefont {Armitage}},\ }\bibfield  {title} {\bibinfo {title} {{Resolving Continua of Fractional Excitations by Spinon Echo in THz 2D Coherent Spectroscopy}},\ }\href {https://doi.org/10.1103/PhysRevLett.122.257401} {\bibfield  {journal} {\bibinfo  {journal} {Phys. Rev. Lett.}\ }\textbf {\bibinfo {volume} {122}},\ \bibinfo {pages} {257401} (\bibinfo {year} {2019})}\BibitemShut {NoStop}%
\bibitem [{\citenamefont {Parameswaran}\ and\ \citenamefont {Gopalakrishnan}(2020)}]{Parameswaran2020}%
  \BibitemOpen
  \bibfield  {author} {\bibinfo {author} {\bibfnamefont {S.~A.}\ \bibnamefont {Parameswaran}}\ and\ \bibinfo {author} {\bibfnamefont {S.}~\bibnamefont {Gopalakrishnan}},\ }\bibfield  {title} {\bibinfo {title} {{Asymptotically Exact Theory for Nonlinear Spectroscopy of Random Quantum Magnets}},\ }\href {https://doi.org/10.1103/PhysRevLett.125.237601} {\bibfield  {journal} {\bibinfo  {journal} {Phys. Rev. Lett.}\ }\textbf {\bibinfo {volume} {125}},\ \bibinfo {pages} {237601} (\bibinfo {year} {2020})}\BibitemShut {NoStop}%
\bibitem [{\citenamefont {Choi}\ \emph {et~al.}(2020)\citenamefont {Choi}, \citenamefont {Lee},\ and\ \citenamefont {Kim}}]{Choi2020}%
  \BibitemOpen
  \bibfield  {author} {\bibinfo {author} {\bibfnamefont {W.}~\bibnamefont {Choi}}, \bibinfo {author} {\bibfnamefont {K.~H.}\ \bibnamefont {Lee}},\ and\ \bibinfo {author} {\bibfnamefont {Y.~B.}\ \bibnamefont {Kim}},\ }\bibfield  {title} {\bibinfo {title} {{Theory of Two-Dimensional Nonlinear Spectroscopy for the Kitaev Spin Liquid}},\ }\href {https://doi.org/10.1103/PhysRevLett.124.117205} {\bibfield  {journal} {\bibinfo  {journal} {Phys. Rev. Lett.}\ }\textbf {\bibinfo {volume} {124}},\ \bibinfo {pages} {117205} (\bibinfo {year} {2020})}\BibitemShut {NoStop}%
\bibitem [{\citenamefont {Li}\ \emph {et~al.}(2021)\citenamefont {Li}, \citenamefont {Oshikawa},\ and\ \citenamefont {Wan}}]{Li2021}%
  \BibitemOpen
  \bibfield  {author} {\bibinfo {author} {\bibfnamefont {Z.-L.}\ \bibnamefont {Li}}, \bibinfo {author} {\bibfnamefont {M.}~\bibnamefont {Oshikawa}},\ and\ \bibinfo {author} {\bibfnamefont {Y.}~\bibnamefont {Wan}},\ }\bibfield  {title} {\bibinfo {title} {{Photon Echo from Lensing of Fractional Excitations in Tomonaga-Luttinger Spin Liquid}},\ }\href {https://doi.org/10.1103/PhysRevX.11.031035} {\bibfield  {journal} {\bibinfo  {journal} {Phys. Rev. X}\ }\textbf {\bibinfo {volume} {11}},\ \bibinfo {pages} {031035} (\bibinfo {year} {2021})}\BibitemShut {NoStop}%
\bibitem [{\citenamefont {Fava}\ \emph {et~al.}(2021)\citenamefont {Fava}, \citenamefont {Biswas}, \citenamefont {Gopalakrishnan}, \citenamefont {Vasseur},\ and\ \citenamefont {Parameswaran}}]{Fava2021}%
  \BibitemOpen
  \bibfield  {author} {\bibinfo {author} {\bibfnamefont {M.}~\bibnamefont {Fava}}, \bibinfo {author} {\bibfnamefont {S.}~\bibnamefont {Biswas}}, \bibinfo {author} {\bibfnamefont {S.}~\bibnamefont {Gopalakrishnan}}, \bibinfo {author} {\bibfnamefont {R.}~\bibnamefont {Vasseur}},\ and\ \bibinfo {author} {\bibfnamefont {S.}~\bibnamefont {Parameswaran}},\ }\bibfield  {title} {\bibinfo {title} {{Hydrodynamic nonlinear response of interacting integrable systems}},\ }\href {https://doi.org/10.1073/pnas.2106945118} {\bibfield  {journal} {\bibinfo  {journal} {Proceedings of the National Academy of Sciences}\ }\textbf {\bibinfo {volume} {118}},\ \bibinfo {pages} {e2106945118} (\bibinfo {year} {2021})}\BibitemShut {NoStop}%
\bibitem [{\citenamefont {Nandkishore}\ \emph {et~al.}(2021)\citenamefont {Nandkishore}, \citenamefont {Choi},\ and\ \citenamefont {Kim}}]{Nandkishore2021}%
  \BibitemOpen
  \bibfield  {author} {\bibinfo {author} {\bibfnamefont {R.~M.}\ \bibnamefont {Nandkishore}}, \bibinfo {author} {\bibfnamefont {W.}~\bibnamefont {Choi}},\ and\ \bibinfo {author} {\bibfnamefont {Y.~B.}\ \bibnamefont {Kim}},\ }\bibfield  {title} {\bibinfo {title} {{Spectroscopic fingerprints of gapped quantum spin liquids, both conventional and fractonic}},\ }\href {https://doi.org/10.1103/PhysRevResearch.3.013254} {\bibfield  {journal} {\bibinfo  {journal} {Phys. Rev. Res.}\ }\textbf {\bibinfo {volume} {3}},\ \bibinfo {pages} {013254} (\bibinfo {year} {2021})}\BibitemShut {NoStop}%
\bibitem [{\citenamefont {Negahdari}\ and\ \citenamefont {Langari}(2023)}]{Negahdari2023}%
  \BibitemOpen
  \bibfield  {author} {\bibinfo {author} {\bibfnamefont {M.~K.}\ \bibnamefont {Negahdari}}\ and\ \bibinfo {author} {\bibfnamefont {A.}~\bibnamefont {Langari}},\ }\bibfield  {title} {\bibinfo {title} {{Nonlinear response of the Kitaev honeycomb lattice model in a weak magnetic field}},\ }\href {https://doi.org/10.1103/PhysRevB.107.134404} {\bibfield  {journal} {\bibinfo  {journal} {Phys. Rev. B}\ }\textbf {\bibinfo {volume} {107}},\ \bibinfo {pages} {134404} (\bibinfo {year} {2023})}\BibitemShut {NoStop}%
\bibitem [{\citenamefont {Hart}\ and\ \citenamefont {Nandkishore}(2023)}]{Hart2023}%
  \BibitemOpen
  \bibfield  {author} {\bibinfo {author} {\bibfnamefont {O.}~\bibnamefont {Hart}}\ and\ \bibinfo {author} {\bibfnamefont {R.}~\bibnamefont {Nandkishore}},\ }\bibfield  {title} {\bibinfo {title} {{Extracting spinon self-energies from two-dimensional coherent spectroscopy}},\ }\href {https://doi.org/10.1103/PhysRevB.107.205143} {\bibfield  {journal} {\bibinfo  {journal} {Phys. Rev. B}\ }\textbf {\bibinfo {volume} {107}},\ \bibinfo {pages} {205143} (\bibinfo {year} {2023})}\BibitemShut {NoStop}%
\bibitem [{\citenamefont {Fava}\ \emph {et~al.}(2023)\citenamefont {Fava}, \citenamefont {Gopalakrishnan}, \citenamefont {Vasseur}, \citenamefont {Essler},\ and\ \citenamefont {Parameswaran}}]{Fava2023}%
  \BibitemOpen
  \bibfield  {author} {\bibinfo {author} {\bibfnamefont {M.}~\bibnamefont {Fava}}, \bibinfo {author} {\bibfnamefont {S.}~\bibnamefont {Gopalakrishnan}}, \bibinfo {author} {\bibfnamefont {R.}~\bibnamefont {Vasseur}}, \bibinfo {author} {\bibfnamefont {F.}~\bibnamefont {Essler}},\ and\ \bibinfo {author} {\bibfnamefont {S.~A.}\ \bibnamefont {Parameswaran}},\ }\bibfield  {title} {\bibinfo {title} {{Divergent Nonlinear Response from Quasiparticle Interactions}},\ }\href {https://doi.org/10.1103/PhysRevLett.131.256505} {\bibfield  {journal} {\bibinfo  {journal} {Phys. Rev. Lett.}\ }\textbf {\bibinfo {volume} {131}},\ \bibinfo {pages} {256505} (\bibinfo {year} {2023})}\BibitemShut {NoStop}%
\bibitem [{\citenamefont {Qiang}\ \emph {et~al.}(2023)\citenamefont {Qiang}, \citenamefont {Quito}, \citenamefont {Trevisan},\ and\ \citenamefont {Orth}}]{Qiang2023}%
  \BibitemOpen
  \bibfield  {author} {\bibinfo {author} {\bibfnamefont {Y.}~\bibnamefont {Qiang}}, \bibinfo {author} {\bibfnamefont {V.~L.}\ \bibnamefont {Quito}}, \bibinfo {author} {\bibfnamefont {T.~V.}\ \bibnamefont {Trevisan}},\ and\ \bibinfo {author} {\bibfnamefont {P.~P.}\ \bibnamefont {Orth}},\ }\href@noop {} {\bibinfo {title} {{Probing Majorana wavefunctions in Kitaev honeycomb spin liquids with second-order two-dimensional spectroscopy}}} (\bibinfo {year} {2023}),\ \Eprint {https://arxiv.org/abs/2301.11243} {arXiv:2301.11243} \BibitemShut {NoStop}%
\bibitem [{\citenamefont {Gao}\ \emph {et~al.}(2023)\citenamefont {Gao}, \citenamefont {Liu}, \citenamefont {Liao},\ and\ \citenamefont {Wan}}]{Gao2023}%
  \BibitemOpen
  \bibfield  {author} {\bibinfo {author} {\bibfnamefont {Q.}~\bibnamefont {Gao}}, \bibinfo {author} {\bibfnamefont {Y.}~\bibnamefont {Liu}}, \bibinfo {author} {\bibfnamefont {H.}~\bibnamefont {Liao}},\ and\ \bibinfo {author} {\bibfnamefont {Y.}~\bibnamefont {Wan}},\ }\bibfield  {title} {\bibinfo {title} {{Two-dimensional coherent spectrum of interacting spinons from matrix product states}},\ }\href {https://doi.org/10.1103/PhysRevB.107.165121} {\bibfield  {journal} {\bibinfo  {journal} {Phys. Rev. B}\ }\textbf {\bibinfo {volume} {107}},\ \bibinfo {pages} {165121} (\bibinfo {year} {2023})}\BibitemShut {NoStop}%
\bibitem [{\citenamefont {Sim}\ \emph {et~al.}(2023{\natexlab{a}})\citenamefont {Sim}, \citenamefont {Knolle},\ and\ \citenamefont {Pollmann}}]{Sim2023}%
  \BibitemOpen
  \bibfield  {author} {\bibinfo {author} {\bibfnamefont {G.}~\bibnamefont {Sim}}, \bibinfo {author} {\bibfnamefont {J.}~\bibnamefont {Knolle}},\ and\ \bibinfo {author} {\bibfnamefont {F.}~\bibnamefont {Pollmann}},\ }\bibfield  {title} {\bibinfo {title} {{Nonlinear spectroscopy of bound states in perturbed Ising spin chains}},\ }\href {https://doi.org/10.1103/PhysRevB.107.L100404} {\bibfield  {journal} {\bibinfo  {journal} {Phys. Rev. B}\ }\textbf {\bibinfo {volume} {107}},\ \bibinfo {pages} {L100404} (\bibinfo {year} {2023}{\natexlab{a}})}\BibitemShut {NoStop}%
\bibitem [{\citenamefont {Sim}\ \emph {et~al.}(2023{\natexlab{b}})\citenamefont {Sim}, \citenamefont {Pollmann},\ and\ \citenamefont {Knolle}}]{Sim2023_2}%
  \BibitemOpen
  \bibfield  {author} {\bibinfo {author} {\bibfnamefont {G.}~\bibnamefont {Sim}}, \bibinfo {author} {\bibfnamefont {F.}~\bibnamefont {Pollmann}},\ and\ \bibinfo {author} {\bibfnamefont {J.}~\bibnamefont {Knolle}},\ }\bibfield  {title} {\bibinfo {title} {{Microscopic details of two-dimensional spectroscopy of one-dimensional quantum Ising magnets}},\ }\href {https://doi.org/10.1103/PhysRevB.108.134423} {\bibfield  {journal} {\bibinfo  {journal} {Phys. Rev. B}\ }\textbf {\bibinfo {volume} {108}},\ \bibinfo {pages} {134423} (\bibinfo {year} {2023}{\natexlab{b}})}\BibitemShut {NoStop}%
\bibitem [{\citenamefont {Zhang}\ \emph {et~al.}(2024{\natexlab{a}})\citenamefont {Zhang}, \citenamefont {Gao}, \citenamefont {Chien}, \citenamefont {Liu}, \citenamefont {Curtis}, \citenamefont {Sung}, \citenamefont {Ma}, \citenamefont {Ren}, \citenamefont {Cao}, \citenamefont {Narang}, \citenamefont {von Hoegen}, \citenamefont {Baldini},\ and\ \citenamefont {Nelson}}]{Zhang2024}%
  \BibitemOpen
  \bibfield  {author} {\bibinfo {author} {\bibfnamefont {Z.}~\bibnamefont {Zhang}}, \bibinfo {author} {\bibfnamefont {F.~Y.}\ \bibnamefont {Gao}}, \bibinfo {author} {\bibfnamefont {Y.-C.}\ \bibnamefont {Chien}}, \bibinfo {author} {\bibfnamefont {Z.-J.}\ \bibnamefont {Liu}}, \bibinfo {author} {\bibfnamefont {J.~B.}\ \bibnamefont {Curtis}}, \bibinfo {author} {\bibfnamefont {E.~R.}\ \bibnamefont {Sung}}, \bibinfo {author} {\bibfnamefont {X.}~\bibnamefont {Ma}}, \bibinfo {author} {\bibfnamefont {W.}~\bibnamefont {Ren}}, \bibinfo {author} {\bibfnamefont {S.}~\bibnamefont {Cao}}, \bibinfo {author} {\bibfnamefont {P.}~\bibnamefont {Narang}}, \bibinfo {author} {\bibfnamefont {A.}~\bibnamefont {von Hoegen}}, \bibinfo {author} {\bibfnamefont {E.}~\bibnamefont {Baldini}},\ and\ \bibinfo {author} {\bibfnamefont {K.~A.}\ \bibnamefont {Nelson}},\ }\bibfield  {title} {\bibinfo {title} {Terahertz-field-driven magnon upconversion in an antiferromagnet},\ }\href {https://doi.org/10.1038/s41567-023-02350-7} {\bibfield  {journal} {\bibinfo  {journal} {Nature Physics}\ } (\bibinfo {year} {2024}{\natexlab{a}})}\BibitemShut {NoStop}%
\bibitem [{\citenamefont {Zhang}\ \emph {et~al.}(2024{\natexlab{b}})\citenamefont {Zhang}, \citenamefont {Gao}, \citenamefont {Curtis}, \citenamefont {Liu}, \citenamefont {Chien}, \citenamefont {von Hoegen}, \citenamefont {Wong}, \citenamefont {Kurihara}, \citenamefont {Suemoto}, \citenamefont {Narang}, \citenamefont {Baldini},\ and\ \citenamefont {Nelson}}]{Zhang2024-2}%
  \BibitemOpen
  \bibfield  {author} {\bibinfo {author} {\bibfnamefont {Z.}~\bibnamefont {Zhang}}, \bibinfo {author} {\bibfnamefont {F.~Y.}\ \bibnamefont {Gao}}, \bibinfo {author} {\bibfnamefont {J.~B.}\ \bibnamefont {Curtis}}, \bibinfo {author} {\bibfnamefont {Z.-J.}\ \bibnamefont {Liu}}, \bibinfo {author} {\bibfnamefont {Y.-C.}\ \bibnamefont {Chien}}, \bibinfo {author} {\bibfnamefont {A.}~\bibnamefont {von Hoegen}}, \bibinfo {author} {\bibfnamefont {M.~T.}\ \bibnamefont {Wong}}, \bibinfo {author} {\bibfnamefont {T.}~\bibnamefont {Kurihara}}, \bibinfo {author} {\bibfnamefont {T.}~\bibnamefont {Suemoto}}, \bibinfo {author} {\bibfnamefont {P.}~\bibnamefont {Narang}}, \bibinfo {author} {\bibfnamefont {E.}~\bibnamefont {Baldini}},\ and\ \bibinfo {author} {\bibfnamefont {K.~A.}\ \bibnamefont {Nelson}},\ }\bibfield  {title} {\bibinfo {title} {Terahertz field-induced nonlinear coupling of two magnon modes in an antiferromagnet},\ }\bibfield  {journal} {\bibinfo  {journal} {Nature Physics}\ }\href {https://doi.org/10.1038/s41567-024-02386-3} {10.1038/s41567-024-02386-3} (\bibinfo {year} {2024}{\natexlab{b}})\BibitemShut {NoStop}%
\bibitem [{\citenamefont {Potts}\ \emph {et~al.}(2024)\citenamefont {Potts}, \citenamefont {Moessner},\ and\ \citenamefont {Benton}}]{Potts2024}%
  \BibitemOpen
  \bibfield  {author} {\bibinfo {author} {\bibfnamefont {M.}~\bibnamefont {Potts}}, \bibinfo {author} {\bibfnamefont {R.}~\bibnamefont {Moessner}},\ and\ \bibinfo {author} {\bibfnamefont {O.}~\bibnamefont {Benton}},\ }\bibfield  {title} {\bibinfo {title} {{Exploiting polarization dependence in two-dimensional coherent spectroscopy: Examples of Ce$_2$Zr$_2$O$_7$ and Nd$_2$Zr$_2$O$_7$}},\ }\href {https://doi.org/10.1103/PhysRevB.109.104435} {\bibfield  {journal} {\bibinfo  {journal} {Phys. Rev. B}\ }\textbf {\bibinfo {volume} {109}},\ \bibinfo {pages} {104435} (\bibinfo {year} {2024})}\BibitemShut {NoStop}%
\bibitem [{\citenamefont {McGinley}\ \emph {et~al.}(2024)\citenamefont {McGinley}, \citenamefont {Fava},\ and\ \citenamefont {Parameswaran}}]{McGinley2024}%
  \BibitemOpen
  \bibfield  {author} {\bibinfo {author} {\bibfnamefont {M.}~\bibnamefont {McGinley}}, \bibinfo {author} {\bibfnamefont {M.}~\bibnamefont {Fava}},\ and\ \bibinfo {author} {\bibfnamefont {S.~A.}\ \bibnamefont {Parameswaran}},\ }\bibfield  {title} {\bibinfo {title} {{Signatures of Fractional Statistics in Nonlinear Pump-Probe Spectroscopy}},\ }\href {https://doi.org/10.1103/PhysRevLett.132.066702} {\bibfield  {journal} {\bibinfo  {journal} {Phys. Rev. Lett.}\ }\textbf {\bibinfo {volume} {132}},\ \bibinfo {pages} {066702} (\bibinfo {year} {2024})}\BibitemShut {NoStop}%
\bibitem [{\citenamefont {Huang}\ \emph {et~al.}(2024)\citenamefont {Huang}, \citenamefont {Luo}, \citenamefont {Mootz}, \citenamefont {Shang}, \citenamefont {Man}, \citenamefont {Su}, \citenamefont {Perakis}, \citenamefont {Yao}, \citenamefont {Wu},\ and\ \citenamefont {Wang}}]{Huang2024}%
  \BibitemOpen
  \bibfield  {author} {\bibinfo {author} {\bibfnamefont {C.}~\bibnamefont {Huang}}, \bibinfo {author} {\bibfnamefont {L.}~\bibnamefont {Luo}}, \bibinfo {author} {\bibfnamefont {M.}~\bibnamefont {Mootz}}, \bibinfo {author} {\bibfnamefont {J.}~\bibnamefont {Shang}}, \bibinfo {author} {\bibfnamefont {P.}~\bibnamefont {Man}}, \bibinfo {author} {\bibfnamefont {L.}~\bibnamefont {Su}}, \bibinfo {author} {\bibfnamefont {I.~E.}\ \bibnamefont {Perakis}}, \bibinfo {author} {\bibfnamefont {Y.~X.}\ \bibnamefont {Yao}}, \bibinfo {author} {\bibfnamefont {A.}~\bibnamefont {Wu}},\ and\ \bibinfo {author} {\bibfnamefont {J.}~\bibnamefont {Wang}},\ }\bibfield  {title} {\bibinfo {title} {{Extreme terahertz magnon multiplication induced by resonant magnetic pulse pairs}},\ }\href {https://doi.org/10.1038/s41467-024-47471-6} {\bibfield  {journal} {\bibinfo  {journal} {Nature Communications}\ }\textbf {\bibinfo {volume} {15}},\ \bibinfo {pages} {3214} (\bibinfo {year} {2024})}\BibitemShut {NoStop}%
\bibitem [{\citenamefont {Papanicolaou}\ and\ \citenamefont {Psaltakis}(1987)}]{Papanicolaou1987}%
  \BibitemOpen
  \bibfield  {author} {\bibinfo {author} {\bibfnamefont {N.}~\bibnamefont {Papanicolaou}}\ and\ \bibinfo {author} {\bibfnamefont {G.~C.}\ \bibnamefont {Psaltakis}},\ }\bibfield  {title} {\bibinfo {title} {{Bethe ansatz for two-magnon bound states in anisotropic magnetic chains of arbitrary spin}},\ }\href {https://doi.org/10.1103/PhysRevB.35.342} {\bibfield  {journal} {\bibinfo  {journal} {Phys. Rev. B}\ }\textbf {\bibinfo {volume} {35}},\ \bibinfo {pages} {342} (\bibinfo {year} {1987})}\BibitemShut {NoStop}%
\bibitem [{\citenamefont {Watanabe}\ \emph {et~al.}(2024{\natexlab{a}})\citenamefont {Watanabe}, \citenamefont {Trebst},\ and\ \citenamefont {Hickey}}]{Watanabe2024}%
  \BibitemOpen
  \bibfield  {author} {\bibinfo {author} {\bibfnamefont {Y.}~\bibnamefont {Watanabe}}, \bibinfo {author} {\bibfnamefont {S.}~\bibnamefont {Trebst}},\ and\ \bibinfo {author} {\bibfnamefont {C.}~\bibnamefont {Hickey}},\ }\href@noop {} {\bibinfo {title} {{Exploring Two-dimensional Coherent Spectroscopy with Exact Diagonalization: Spinons and Confinement in 1D Quantum Magnets}}} (\bibinfo {year} {2024}{\natexlab{a}}),\ \Eprint {https://arxiv.org/abs/2401.17266} {arXiv:2401.17266} \BibitemShut {NoStop}%
\bibitem [{\citenamefont {Chauhan}\ \emph {et~al.}(2020)\citenamefont {Chauhan}, \citenamefont {Mahmood}, \citenamefont {Changlani}, \citenamefont {Koohpayeh},\ and\ \citenamefont {Armitage}}]{Chauhan2020}%
  \BibitemOpen
  \bibfield  {author} {\bibinfo {author} {\bibfnamefont {P.}~\bibnamefont {Chauhan}}, \bibinfo {author} {\bibfnamefont {F.}~\bibnamefont {Mahmood}}, \bibinfo {author} {\bibfnamefont {H.~J.}\ \bibnamefont {Changlani}}, \bibinfo {author} {\bibfnamefont {S.~M.}\ \bibnamefont {Koohpayeh}},\ and\ \bibinfo {author} {\bibfnamefont {N.~P.}\ \bibnamefont {Armitage}},\ }\bibfield  {title} {\bibinfo {title} {{Tunable Magnon Interactions in a Ferromagnetic Spin-1 Chain}},\ }\href {https://doi.org/10.1103/PhysRevLett.124.037203} {\bibfield  {journal} {\bibinfo  {journal} {Phys. Rev. Lett.}\ }\textbf {\bibinfo {volume} {124}},\ \bibinfo {pages} {037203} (\bibinfo {year} {2020})}\BibitemShut {NoStop}%
\bibitem [{\citenamefont {Hamm}\ and\ \citenamefont {Zanni}(2011)}]{Hamm_Zanni_2011}%
  \BibitemOpen
  \bibfield  {author} {\bibinfo {author} {\bibfnamefont {P.}~\bibnamefont {Hamm}}\ and\ \bibinfo {author} {\bibfnamefont {M.}~\bibnamefont {Zanni}},\ }\href@noop {} {\emph {\bibinfo {title} {{Concepts and Methods of 2D Infrared Spectroscopy}}}}\ (\bibinfo  {publisher} {Cambridge University Press},\ \bibinfo {year} {2011})\BibitemShut {NoStop}%
\bibitem [{\citenamefont {Koch}\ \emph {et~al.}(2003)\citenamefont {Koch}, \citenamefont {Waldmann}, \citenamefont {M\"uller}, \citenamefont {Reimann},\ and\ \citenamefont {Saalfrank}}]{Koch2003}%
  \BibitemOpen
  \bibfield  {author} {\bibinfo {author} {\bibfnamefont {R.}~\bibnamefont {Koch}}, \bibinfo {author} {\bibfnamefont {O.}~\bibnamefont {Waldmann}}, \bibinfo {author} {\bibfnamefont {P.}~\bibnamefont {M\"uller}}, \bibinfo {author} {\bibfnamefont {U.}~\bibnamefont {Reimann}},\ and\ \bibinfo {author} {\bibfnamefont {R.~W.}\ \bibnamefont {Saalfrank}},\ }\bibfield  {title} {\bibinfo {title} {{Ferromagnetic coupling and magnetic anisotropy in molecular Ni(II) squares}},\ }\href {https://doi.org/10.1103/PhysRevB.67.094407} {\bibfield  {journal} {\bibinfo  {journal} {Phys. Rev. B}\ }\textbf {\bibinfo {volume} {67}},\ \bibinfo {pages} {094407} (\bibinfo {year} {2003})}\BibitemShut {NoStop}%
\bibitem [{\citenamefont {Lee}\ \emph {et~al.}(2012)\citenamefont {Lee}, \citenamefont {Onoda},\ and\ \citenamefont {Balents}}]{Lee2012}%
  \BibitemOpen
  \bibfield  {author} {\bibinfo {author} {\bibfnamefont {S.}~\bibnamefont {Lee}}, \bibinfo {author} {\bibfnamefont {S.}~\bibnamefont {Onoda}},\ and\ \bibinfo {author} {\bibfnamefont {L.}~\bibnamefont {Balents}},\ }\bibfield  {title} {\bibinfo {title} {{Generic quantum spin ice}},\ }\href {https://doi.org/10.1103/PhysRevB.86.104412} {\bibfield  {journal} {\bibinfo  {journal} {Phys. Rev. B}\ }\textbf {\bibinfo {volume} {86}},\ \bibinfo {pages} {104412} (\bibinfo {year} {2012})}\BibitemShut {NoStop}%
\bibitem [{\citenamefont {Liu}\ \emph {et~al.}(2018)\citenamefont {Liu}, \citenamefont {Li},\ and\ \citenamefont {Chen}}]{Liu2018}%
  \BibitemOpen
  \bibfield  {author} {\bibinfo {author} {\bibfnamefont {C.}~\bibnamefont {Liu}}, \bibinfo {author} {\bibfnamefont {Y.-D.}\ \bibnamefont {Li}},\ and\ \bibinfo {author} {\bibfnamefont {G.}~\bibnamefont {Chen}},\ }\bibfield  {title} {\bibinfo {title} {{Selective measurements of intertwined multipolar orders: Non-Kramers doublets on a triangular lattice}},\ }\href {https://doi.org/10.1103/PhysRevB.98.045119} {\bibfield  {journal} {\bibinfo  {journal} {Phys. Rev. B}\ }\textbf {\bibinfo {volume} {98}},\ \bibinfo {pages} {045119} (\bibinfo {year} {2018})}\BibitemShut {NoStop}%
\bibitem [{\citenamefont {Sakai}\ and\ \citenamefont {Nakatsuji}(2011)}]{Sakai2011}%
  \BibitemOpen
  \bibfield  {author} {\bibinfo {author} {\bibfnamefont {A.}~\bibnamefont {Sakai}}\ and\ \bibinfo {author} {\bibfnamefont {S.}~\bibnamefont {Nakatsuji}},\ }\bibfield  {title} {\bibinfo {title} {{Kondo Effects and Multipolar Order in the Cubic PrTr$_2$Al$_{20}$ (Tr=Ti, V)}},\ }\href {https://doi.org/10.1143/JPSJ.80.063701} {\bibfield  {journal} {\bibinfo  {journal} {Journal of the Physical Society of Japan}\ }\textbf {\bibinfo {volume} {80}},\ \bibinfo {pages} {063701} (\bibinfo {year} {2011})}\BibitemShut {NoStop}%
\bibitem [{\citenamefont {Freyer}\ \emph {et~al.}(2018)\citenamefont {Freyer}, \citenamefont {Attig}, \citenamefont {Lee}, \citenamefont {Paramekanti}, \citenamefont {Trebst},\ and\ \citenamefont {Kim}}]{Freyer2018}%
  \BibitemOpen
  \bibfield  {author} {\bibinfo {author} {\bibfnamefont {F.}~\bibnamefont {Freyer}}, \bibinfo {author} {\bibfnamefont {J.}~\bibnamefont {Attig}}, \bibinfo {author} {\bibfnamefont {S.}~\bibnamefont {Lee}}, \bibinfo {author} {\bibfnamefont {A.}~\bibnamefont {Paramekanti}}, \bibinfo {author} {\bibfnamefont {S.}~\bibnamefont {Trebst}},\ and\ \bibinfo {author} {\bibfnamefont {Y.~B.}\ \bibnamefont {Kim}},\ }\bibfield  {title} {\bibinfo {title} {{Two-stage multipolar ordering in Pr$T_2$Al$_{20}$ Kondo materials}},\ }\href {https://doi.org/10.1103/PhysRevB.97.115111} {\bibfield  {journal} {\bibinfo  {journal} {Phys. Rev. B}\ }\textbf {\bibinfo {volume} {97}},\ \bibinfo {pages} {115111} (\bibinfo {year} {2018})}\BibitemShut {NoStop}%
\bibitem [{\citenamefont {Pradhan}\ \emph {et~al.}(2024)\citenamefont {Pradhan}, \citenamefont {Paramekanti},\ and\ \citenamefont {Saha-Dasgupta}}]{Pradhan2024}%
  \BibitemOpen
  \bibfield  {author} {\bibinfo {author} {\bibfnamefont {K.}~\bibnamefont {Pradhan}}, \bibinfo {author} {\bibfnamefont {A.}~\bibnamefont {Paramekanti}},\ and\ \bibinfo {author} {\bibfnamefont {T.}~\bibnamefont {Saha-Dasgupta}},\ }\bibfield  {title} {\bibinfo {title} {{Multipolar magnetism in $5{d}^{2}$ vacancy-ordered halide double perovskites}},\ }\href {https://doi.org/10.1103/PhysRevB.109.184416} {\bibfield  {journal} {\bibinfo  {journal} {Phys. Rev. B}\ }\textbf {\bibinfo {volume} {109}},\ \bibinfo {pages} {184416} (\bibinfo {year} {2024})}\BibitemShut {NoStop}%
\bibitem [{\citenamefont {Zvyagin}\ \emph {et~al.}(2008)\citenamefont {Zvyagin}, \citenamefont {Batista}, \citenamefont {Krzystek}, \citenamefont {Zapf}, \citenamefont {Jaime}, \citenamefont {Paduan-Filho},\ and\ \citenamefont {Wosnitza}}]{Zvyagin2008}%
  \BibitemOpen
  \bibfield  {author} {\bibinfo {author} {\bibfnamefont {S.}~\bibnamefont {Zvyagin}}, \bibinfo {author} {\bibfnamefont {C.}~\bibnamefont {Batista}}, \bibinfo {author} {\bibfnamefont {J.}~\bibnamefont {Krzystek}}, \bibinfo {author} {\bibfnamefont {V.}~\bibnamefont {Zapf}}, \bibinfo {author} {\bibfnamefont {M.}~\bibnamefont {Jaime}}, \bibinfo {author} {\bibfnamefont {A.}~\bibnamefont {Paduan-Filho}},\ and\ \bibinfo {author} {\bibfnamefont {J.}~\bibnamefont {Wosnitza}},\ }\bibfield  {title} {\bibinfo {title} {{Observation of two-magnon bound states in the spin-1 anisotropic Heisenberg antiferromagnetic chain system NiCl$_2$-4SC(NH$_2$)$_2$}},\ }\href {https://doi.org/https://doi.org/10.1016/j.physb.2007.10.174} {\bibfield  {journal} {\bibinfo  {journal} {Physica B: Condensed Matter}\ }\textbf {\bibinfo {volume} {403}},\ \bibinfo {pages} {1497} (\bibinfo {year} {2008})}\BibitemShut {NoStop}%
\bibitem [{\citenamefont {Akaki}\ \emph {et~al.}(2017)\citenamefont {Akaki}, \citenamefont {Yoshizawa}, \citenamefont {Okutani}, \citenamefont {Kida}, \citenamefont {Romh\'anyi}, \citenamefont {Penc},\ and\ \citenamefont {Hagiwara}}]{Akaki2017}%
  \BibitemOpen
  \bibfield  {author} {\bibinfo {author} {\bibfnamefont {M.}~\bibnamefont {Akaki}}, \bibinfo {author} {\bibfnamefont {D.}~\bibnamefont {Yoshizawa}}, \bibinfo {author} {\bibfnamefont {A.}~\bibnamefont {Okutani}}, \bibinfo {author} {\bibfnamefont {T.}~\bibnamefont {Kida}}, \bibinfo {author} {\bibfnamefont {J.}~\bibnamefont {Romh\'anyi}}, \bibinfo {author} {\bibfnamefont {K.}~\bibnamefont {Penc}},\ and\ \bibinfo {author} {\bibfnamefont {M.}~\bibnamefont {Hagiwara}},\ }\bibfield  {title} {\bibinfo {title} {{Direct observation of spin-quadrupolar excitations in Sr$_2$CoGe$_2$O$_7$ by high-field electron spin resonance}},\ }\href {https://doi.org/10.1103/PhysRevB.96.214406} {\bibfield  {journal} {\bibinfo  {journal} {Phys. Rev. B}\ }\textbf {\bibinfo {volume} {96}},\ \bibinfo {pages} {214406} (\bibinfo {year} {2017})}\BibitemShut {NoStop}%
\bibitem [{\citenamefont {Coldea}\ \emph {et~al.}(2010)\citenamefont {Coldea}, \citenamefont {Tennant}, \citenamefont {Wheeler}, \citenamefont {Wawrzynska}, \citenamefont {Prabhakaran}, \citenamefont {Telling}, \citenamefont {Habicht}, \citenamefont {Smeibidl},\ and\ \citenamefont {Kiefer}}]{Coldea2010}%
  \BibitemOpen
  \bibfield  {author} {\bibinfo {author} {\bibfnamefont {R.}~\bibnamefont {Coldea}}, \bibinfo {author} {\bibfnamefont {D.~A.}\ \bibnamefont {Tennant}}, \bibinfo {author} {\bibfnamefont {E.~M.}\ \bibnamefont {Wheeler}}, \bibinfo {author} {\bibfnamefont {E.}~\bibnamefont {Wawrzynska}}, \bibinfo {author} {\bibfnamefont {D.}~\bibnamefont {Prabhakaran}}, \bibinfo {author} {\bibfnamefont {M.}~\bibnamefont {Telling}}, \bibinfo {author} {\bibfnamefont {K.}~\bibnamefont {Habicht}}, \bibinfo {author} {\bibfnamefont {P.}~\bibnamefont {Smeibidl}},\ and\ \bibinfo {author} {\bibfnamefont {K.}~\bibnamefont {Kiefer}},\ }\bibfield  {title} {\bibinfo {title} {{Quantum Criticality in an Ising Chain: Experimental Evidence for Emergent E8 Symmetry}},\ }\href {https://doi.org/10.1126/science.1180085} {\bibfield  {journal} {\bibinfo  {journal} {Science}\ }\textbf {\bibinfo {volume} {327}},\ \bibinfo {pages} {177} (\bibinfo {year} {2010})}\BibitemShut {NoStop}%
\bibitem [{\citenamefont {Nishida}\ \emph {et~al.}(2013)\citenamefont {Nishida}, \citenamefont {Kato},\ and\ \citenamefont {Batista}}]{Nishida2013}%
  \BibitemOpen
  \bibfield  {author} {\bibinfo {author} {\bibfnamefont {Y.}~\bibnamefont {Nishida}}, \bibinfo {author} {\bibfnamefont {Y.}~\bibnamefont {Kato}},\ and\ \bibinfo {author} {\bibfnamefont {C.~D.}\ \bibnamefont {Batista}},\ }\bibfield  {title} {\bibinfo {title} {{Efimov effect in quantum magnets}},\ }\href {https://doi.org/10.1038/nphys2523} {\bibfield  {journal} {\bibinfo  {journal} {Nature Physics}\ }\textbf {\bibinfo {volume} {9}},\ \bibinfo {pages} {93} (\bibinfo {year} {2013})}\BibitemShut {NoStop}%
\bibitem [{\citenamefont {Morris}\ \emph {et~al.}(2014)\citenamefont {Morris}, \citenamefont {Vald\'es~Aguilar}, \citenamefont {Ghosh}, \citenamefont {Koohpayeh}, \citenamefont {Krizan}, \citenamefont {Cava}, \citenamefont {Tchernyshyov}, \citenamefont {McQueen},\ and\ \citenamefont {Armitage}}]{Morris2014}%
  \BibitemOpen
  \bibfield  {author} {\bibinfo {author} {\bibfnamefont {C.~M.}\ \bibnamefont {Morris}}, \bibinfo {author} {\bibfnamefont {R.}~\bibnamefont {Vald\'es~Aguilar}}, \bibinfo {author} {\bibfnamefont {A.}~\bibnamefont {Ghosh}}, \bibinfo {author} {\bibfnamefont {S.~M.}\ \bibnamefont {Koohpayeh}}, \bibinfo {author} {\bibfnamefont {J.}~\bibnamefont {Krizan}}, \bibinfo {author} {\bibfnamefont {R.~J.}\ \bibnamefont {Cava}}, \bibinfo {author} {\bibfnamefont {O.}~\bibnamefont {Tchernyshyov}}, \bibinfo {author} {\bibfnamefont {T.~M.}\ \bibnamefont {McQueen}},\ and\ \bibinfo {author} {\bibfnamefont {N.~P.}\ \bibnamefont {Armitage}},\ }\bibfield  {title} {\bibinfo {title} {{Hierarchy of Bound States in the One-Dimensional Ferromagnetic Ising Chain CoNb$_2$O$_6$ Investigated by High-Resolution Time-Domain Terahertz Spectroscopy}},\ }\href {https://doi.org/10.1103/PhysRevLett.112.137403} {\bibfield  {journal} {\bibinfo  {journal} {Phys. Rev. Lett.}\ }\textbf {\bibinfo {volume} {112}},\ \bibinfo {pages} {137403} (\bibinfo {year} {2014})}\BibitemShut {NoStop}%
\bibitem [{\citenamefont {Grenier}\ \emph {et~al.}(2015)\citenamefont {Grenier}, \citenamefont {Petit}, \citenamefont {Simonet}, \citenamefont {Can\'evet}, \citenamefont {Regnault}, \citenamefont {Raymond}, \citenamefont {Canals}, \citenamefont {Berthier},\ and\ \citenamefont {Lejay}}]{Grenier2015}%
  \BibitemOpen
  \bibfield  {author} {\bibinfo {author} {\bibfnamefont {B.}~\bibnamefont {Grenier}}, \bibinfo {author} {\bibfnamefont {S.}~\bibnamefont {Petit}}, \bibinfo {author} {\bibfnamefont {V.}~\bibnamefont {Simonet}}, \bibinfo {author} {\bibfnamefont {E.}~\bibnamefont {Can\'evet}}, \bibinfo {author} {\bibfnamefont {L.-P.}\ \bibnamefont {Regnault}}, \bibinfo {author} {\bibfnamefont {S.}~\bibnamefont {Raymond}}, \bibinfo {author} {\bibfnamefont {B.}~\bibnamefont {Canals}}, \bibinfo {author} {\bibfnamefont {C.}~\bibnamefont {Berthier}},\ and\ \bibinfo {author} {\bibfnamefont {P.}~\bibnamefont {Lejay}},\ }\bibfield  {title} {\bibinfo {title} {{Longitudinal and Transverse Zeeman Ladders in the Ising-Like Chain Antiferromagnet BaCo$_2$V$_2$O$_8$}},\ }\href {https://doi.org/10.1103/PhysRevLett.114.017201} {\bibfield  {journal} {\bibinfo  {journal} {Phys. Rev. Lett.}\ }\textbf {\bibinfo {volume} {114}},\ \bibinfo {pages} {017201} (\bibinfo {year} {2015})}\BibitemShut {NoStop}%
\bibitem [{\citenamefont {Wang}\ \emph {et~al.}(2018)\citenamefont {Wang}, \citenamefont {Wu}, \citenamefont {Yang}, \citenamefont {Bera}, \citenamefont {Kamenskyi}, \citenamefont {Islam}, \citenamefont {Xu}, \citenamefont {Law}, \citenamefont {Lake}, \citenamefont {Wu},\ and\ \citenamefont {Loidl}}]{Wang2018}%
  \BibitemOpen
  \bibfield  {author} {\bibinfo {author} {\bibfnamefont {Z.}~\bibnamefont {Wang}}, \bibinfo {author} {\bibfnamefont {J.}~\bibnamefont {Wu}}, \bibinfo {author} {\bibfnamefont {W.}~\bibnamefont {Yang}}, \bibinfo {author} {\bibfnamefont {A.~K.}\ \bibnamefont {Bera}}, \bibinfo {author} {\bibfnamefont {D.}~\bibnamefont {Kamenskyi}}, \bibinfo {author} {\bibfnamefont {A.~T. M.~N.}\ \bibnamefont {Islam}}, \bibinfo {author} {\bibfnamefont {S.}~\bibnamefont {Xu}}, \bibinfo {author} {\bibfnamefont {J.~M.}\ \bibnamefont {Law}}, \bibinfo {author} {\bibfnamefont {B.}~\bibnamefont {Lake}}, \bibinfo {author} {\bibfnamefont {C.}~\bibnamefont {Wu}},\ and\ \bibinfo {author} {\bibfnamefont {A.}~\bibnamefont {Loidl}},\ }\bibfield  {title} {\bibinfo {title} {{Experimental observation of Bethe strings}},\ }\href {https://doi.org/10.1038/nature25466} {\bibfield  {journal} {\bibinfo  {journal} {Nature}\ }\textbf {\bibinfo {volume} {554}},\ \bibinfo {pages} {219} (\bibinfo {year} {2018})}\BibitemShut {NoStop}%
\bibitem [{\citenamefont {Dally}\ \emph {et~al.}(2020)\citenamefont {Dally}, \citenamefont {Heng}, \citenamefont {Keselman}, \citenamefont {Bordelon}, \citenamefont {Stone}, \citenamefont {Balents},\ and\ \citenamefont {Wilson}}]{Dally2020}%
  \BibitemOpen
  \bibfield  {author} {\bibinfo {author} {\bibfnamefont {R.~L.}\ \bibnamefont {Dally}}, \bibinfo {author} {\bibfnamefont {A.~J.~R.}\ \bibnamefont {Heng}}, \bibinfo {author} {\bibfnamefont {A.}~\bibnamefont {Keselman}}, \bibinfo {author} {\bibfnamefont {M.~M.}\ \bibnamefont {Bordelon}}, \bibinfo {author} {\bibfnamefont {M.~B.}\ \bibnamefont {Stone}}, \bibinfo {author} {\bibfnamefont {L.}~\bibnamefont {Balents}},\ and\ \bibinfo {author} {\bibfnamefont {S.~D.}\ \bibnamefont {Wilson}},\ }\bibfield  {title} {\bibinfo {title} {{Three-Magnon Bound State in the Quasi-One-Dimensional Antiferromagnet $\alpha$-NaMnO$_2$}},\ }\href {https://doi.org/10.1103/PhysRevLett.124.197203} {\bibfield  {journal} {\bibinfo  {journal} {Phys. Rev. Lett.}\ }\textbf {\bibinfo {volume} {124}},\ \bibinfo {pages} {197203} (\bibinfo {year} {2020})}\BibitemShut {NoStop}%
\bibitem [{\citenamefont {Watanabe}\ \emph {et~al.}(2024{\natexlab{b}})\citenamefont {Watanabe}, \citenamefont {Trebst},\ and\ \citenamefont {Hickey}}]{zenodo_repository}%
  \BibitemOpen
  \bibfield  {author} {\bibinfo {author} {\bibfnamefont {Y.}~\bibnamefont {Watanabe}}, \bibinfo {author} {\bibfnamefont {S.}~\bibnamefont {Trebst}},\ and\ \bibinfo {author} {\bibfnamefont {C.}~\bibnamefont {Hickey}},\ }\bibfield  {title} {\bibinfo {title} {{Data underpinning "Revealing Quadrupolar Excitations with Non-Linear Spectroscopy"}},\ }\href {https://doi.org/10.5281/zenodo.11106522} {10.5281/zenodo.11106522} (\bibinfo {year} {2024}{\natexlab{b}})\BibitemShut {NoStop}%
\end{thebibliography}%


\clearpage



\onecolumngrid
\section*{Supplementary Material}


\makeatletter
\renewcommand \thesection{S\@arabic\c@section}
\renewcommand\thetable{S\@arabic\c@table}
\renewcommand \thefigure{S\@arabic\c@figure}
\renewcommand \theequation{S\@arabic\c@equation}
\makeatother
\setcounter{figure}{0}
\setcounter{equation}{0}

\section{Dipolar and Quadrupolar Operators}
In a spin-1 system, local moments can possess both dipolar and quadrupolar character. Therefore, it is useful to define both the dipolar and quadrupolar operators in terms of the spin-1 operators $S_i^\alpha$ ($\alpha = x, y, z$). Taking the $z$-axis as the spin quantization axis, the dipolar and quadrupolar operators are given by:
\begin{align}
    \begin{split}
      \textrm{Dipolar:} \quad & S_i^x = \frac{1}{2}(S_i^+ + S_i^-)\,,\quad S_i^y = -\frac{i}{2}(S_i^+ - S_i^-)\,,\quad S_i^z. \\
      \textrm{Quadrupolar:} \quad & Q_i^{x^2-y^2} = \frac{1}{2} (S_i^+ S_i^+ + S_i^- S_i^-)\,,\quad Q_i^{xy} = -\frac{i}{2} (S_i^+ S_i^+ - S_i^- S_i^-)\,, \\
      & Q_i^{xz} = S_i^z (S_i^+ + S_i^-)\,,\quad Q_i^{yz} = -iS_i^z (S_i^+ - S_i^-)\,,\quad Q_i^{3z^2 - r^2} = \sqrt{3} S_i^z S_i^z - \frac{2}{\sqrt{3}}.
  \end{split}
\end{align}
The dipolar operators $S_i^x$ and $S_i^y$, as well as the quadrupolar operators $Q_i^{xz}$ and $Q_i^{yz}$, produce $|\Delta S_i^z| =1$ transitions. On the other hand, the quadrupolar operators $Q_i^{x^2-y^2}$ and $Q_i^{xy}$ produce $|\Delta S_i^z| =2$ transitions. The remaining operators, $S_i^z$ and $Q_i^{3z^2-r^2}$, do not change $S_i^z$, i.e.~$|\Delta S_i^z| =0$.


\section{Higher-dimensional FM spin-1 systems}
\begin{figure}[h]
  \centering
  \includegraphics[width=1.0\textwidth]{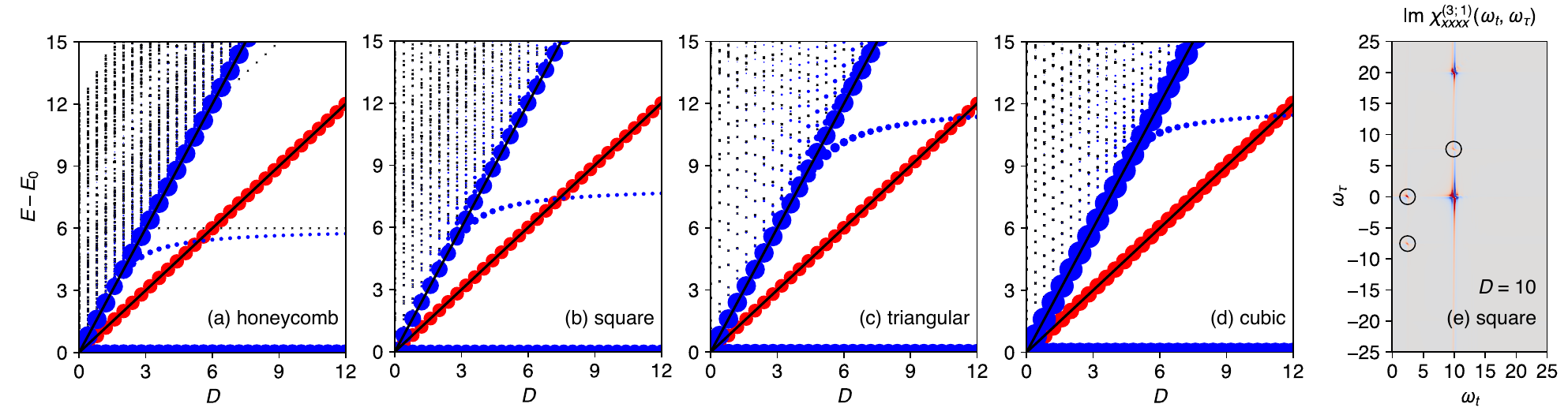}
  \caption{{\bf Quadrupolar excitation in higher spatial dimensions}. Excitation spectrums of spin-1 FM Heisenberg model on (a) honeycomb ($L=20$), (b) square ($L=20$), (c) triangular ($L=20$), and (d) cubic ($L=12$) lattice. The same color scheme is used as in Fig.~\ref{fig:chi1_Ddep}. The excitation energy marked by black cross indicates that it has neither dipolar nor quadrupolar transition amplitude. (e) $\chi^{(3;1)}_{xxxx}(\omega_t, \omega_\tau)$ for the square lattice.}
  \label{fig:higher_dimension}
\end{figure}

In the main text, we have shown numerical results for the one-dimensional Heisenberg spin-1 chain. However, all of the results are qualitatively similar in higher-dimensions as well. 
Figure~\ref{fig:higher_dimension} shows the excitation spectrum of the spin-1 FM Heisenberg model with single-ion anisotropy on a range of 2-d and 3-d lattices obtained by ED. In all cases, the 1-magnon energy $\omega_{1m} = D$.
At large $D$, the energy of the SIBS is given by $zJ$, where $z$ is the coordination number of the lattice. As $D$ decreases, the energy of the SIBS decreases and eventually merges with the 2-magnon continuum.
As a result, they all share qualitatively similar 2DCS spectra, as shown in Fig.~\ref{fig:higher_dimension}(e) for the square lattice case as an example. There, the 2DCS spectrum exhibits three distinct peaks associated with the SIBS, just as in the 1-d case presented in the main text.

Note that, in the honeycomb lattice, due to the two-site unit cell, there are two SIBS modes, with the additional mode having a fixed energy of $zJ$. This additional mode, however, is not active in the 2DCS response due to the sublattice effect~\cite{Legros2021}. 

\section{Impact of XXZ anisotropy}
\begin{figure}[h]
  \centering
  \includegraphics[width=0.9\textwidth]{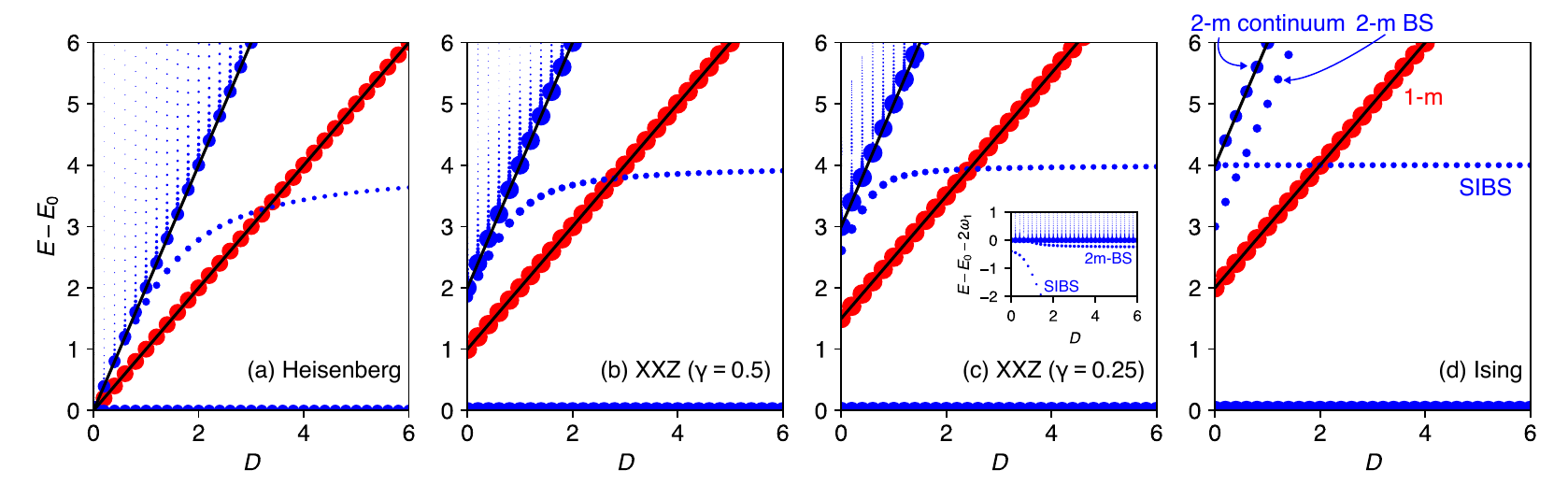}
  \caption{{\bf Quadrupolar excitation in XXZ FM chain}. Excitation spectrums of spin-1 FM chains with XXZ anisotropy (Eq.~\ref{eq:XXZ}). (a) Heisenberg ($\gamma = 1$), (b) XXZ ($\gamma = 0.5$), (c) XXZ ($\gamma = 0.25$), and (d) Ising ($\gamma = 0$) for $L=100$. The same color scheme is used as in Fig.~\ref{fig:chi1_Ddep}.}
  \label{fig:xxz}
\end{figure}

In the main text we focused on the Heisenberg limit of the model. Here we consider the impact of XXZ anisotropy on the spectrum. The Hamiltonian is now given by
\begin{equation}
  \mathcal{H} = -J\sum_i\left[\gamma(S^x_iS^x_{i+1} + S^y_iS^y_{i+1}) + S^z_iS^z_{i+1} \right] - D\sum_i(S^z_i)^2,
  \label{eq:XXZ}
\end{equation}
where $0 < \gamma < 1$. Figure~\ref{fig:xxz} shows the excitation spectrum of the spin-1 FM chain with varying $\gamma$. In the Ising limit there is a 2-magnon bound state (2m-BS) where the energy is $J$ lower than the lower limit of the 2-magnon continuum. As $\gamma$ increases, the 2m-BS state quickly dissolves into the continuum, e.g. for $\gamma = 0.25$ the energy of the 2m-BS is already close to the lower edge of the continuum, as shown in the inset of Fig.~\ref{fig:xxz}(c), and enters the continuum below $D/J < 0.8$. The SIBS, on the other hand, remains relatively well separated from the 2-magnon continuum for all $\gamma$ values as long as $D/J$ is large enough.

\section{Localized nature of SIBS}
\begin{figure}[h]
  \centering
  \includegraphics[width=0.6\textwidth]{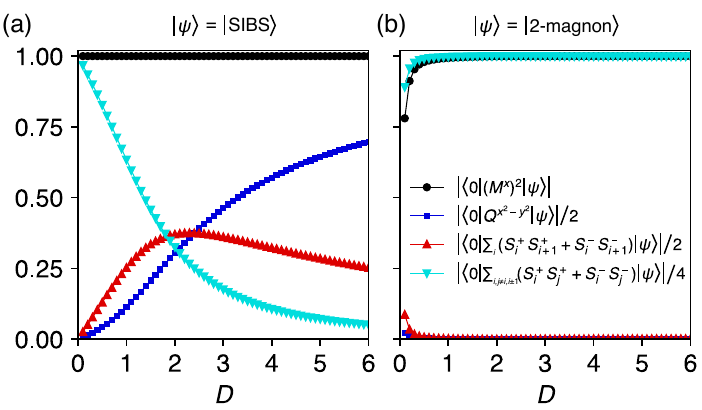}
  \caption{{\bf Localized nature of SIBS.} Transition amplitude $\left|\braket{0|(M^x)^2|\psi}\right|$ and first, second and third terms in Eq.~(\ref{eq:Mx2}). (a) $\ket{\psi}$ is SIBS, (b) $\ket{\psi}$ is lowest energy state of 2-magnon continuum. The transition amplitude is calculated by ED for $L=100$. Values are normalized by the sum of the absolute value of all individual terms.} \label{fig:mix}
\end{figure}

As discussed in the main text, though the SIBS and 2-magnon states can mix, they have a distinct underlying nature at large enough $D/J$. To elucidate this underlying nature more concretely, we calculate the magnitude of the individual terms in the transition amplitude $\braket{\psi|(M^x)^2|0}$ [Eq.~(\ref{eq:Mx2})] for $\ket{\psi}$ being the SIBS and compare the result with that for the lowest energy state of the 2-magnon continuum. As seen in Fig.~\ref{fig:mix} for the 1-d chain case, for $D/J\gtrsim 2.5$ the SIBS is dominated by the purely on-site quadrupolar contribution, $Q_i^{x^2-y^2}$, while the 2-magnon state is always fully dominated by the 2-magnon contribution, $(S_i^+S_j^+ + S_i^-S_j^-)$. For smaller $D/J$, the SIBS mixes with the 2-magnon continuum and, for $D/J\lesssim 1.5$, it is actually dominated by 2-magnon contributions and loses its distinct SIBS nature.  

\section{Hybridized SIBS}
In the main text, we have argued that the peaks at $(\omega_t, \omega_\tau) = (\omega_2 - \omega_1, \pm \omega_1)$ and $(\omega_2 - \omega_1, \pm \omega_2)$ in $\chi^{(3;2)}_{xxxx}(\omega_t, \omega_\tau) \equiv \text{FT}[\chi^{(3)}_{xxxx}(t,0,\tau)]$ can be used to reveal the quadrupolar nature of the hybridized SIBS. Here we provide a detailed derivation of this result.

We will focus on the peaks at $\omega_t = \omega_2 - \omega_1$. $\chi^{(3;2)}_{xxxx}$ can be written in the time domain as
\begin{equation}
\chi^{(3)}_{xxxx}(t,0,\tau) = -\frac{1}{N}\sum_{PQR} m^x_{0R}m^x_{RQ} m^x_{QP} m^x_{P0} \left[\sin\left( E_P \tau  + \Delta E_{PQ} t \right) +  2\sin\left( -E_R \tau  + \Delta E_{PQ} t \right) +   \sin\left( E_P \tau  + E_R t \right)  \right].
\end{equation}
We assume that a single $M^x$ operator excites either the first excited state $\psi_1$ or the second excited state $\psi_2$, and the transition probability to all higher energy states is negligible. The third term in the above equation cannot contribute to the peak as the energy of the state $\ket{R}$ created by a single $M^x$ operator is larger than $\omega_2 - \omega_1$. Setting $|\Delta E_{PQ}| = \omega_2 - \omega_1$ requires $(P, Q)$ to be $(\psi_1, \psi_2)$ or $(\psi_2, \psi_1)$. Furthermore, with the field pulses aligned along the $x$-axis, $m^x_{00} = m^x_{11} = m^x_{22} = 0$. Therefore, the possible processes are
\begin{itemize}
  \item $P = \psi_1$, $Q = \psi_2$, $R = \psi_1$ $\rightarrow$ $\omega_\tau = \omega_1$ : $-2\left|m^x_{01}\right|^2\left|m^x_{12}\right|^2$, $\omega_\tau = -\omega_1$ : $-\left|m^x_{01}\right|^2\left|m^x_{12}\right|^2$
  \item $P = \psi_2$, $Q = \psi_1$, $R = \psi_2$ $\rightarrow$ $\omega_\tau = \omega_2$ : $\left|m^x_{02}\right|^2\left|m^x_{12}\right|^2$, $\omega_\tau = -\omega_2$ : $2\left|m^x_{02}\right|^2\left|m^x_{12}\right|^2$
\end{itemize}
In that case, the peak intensities at $\omega_t = \omega_2 - \omega_1$ are given by
\begin{equation}
  \begin{split}
    I_{(\omega_2 - \omega_1, \omega_2)} & \propto \left|m^x_{02}\right|^2\left|m^x_{12}\right|^2,\, 
    I_{(\omega_2 - \omega_1, -\omega_2)}  \propto 2\left|m^x_{02}\right|^2\left|m^x_{12}\right|^2 \\
    I_{(\omega_2 - \omega_1, \omega_1)} & \propto -2\left|m^x_{01}\right|^2\left|m^x_{12}\right|^2,\,
    I_{(\omega_2 - \omega_1, -\omega_1)}  \propto -\left|m^x_{01}\right|^2\left|m^x_{12}\right|^2.
  \end{split}
\end{equation}
On the other hand, the quadrupolar transition amplitudes are given by
\begin{equation}
  \begin{split}
   \left|\braket{\psi_2|(M^x)^2|0}\right|^2 & = \sum_{n,m}\braket{0|M^x|n}\braket{n|M^x|\psi_2}\braket{\psi_2|M^x|m}\braket{m|M^x|0} \\
   & \approx \sum_{n,m = 0,\psi_1,\psi_2}\braket{0|M^x|n}\braket{n|M^x|\psi_2}\braket{\psi_2|M^x|m}\braket{m|M^x|0} \\
   & = \braket{0|M^x|\psi_1}\braket{\psi_1|M^x|\psi_2}\braket{\psi_2|M^x|\psi_1}\braket{\psi_1|M^x|0} \, (\because m^x_{00} = m^x_{11} = m^x_{22} = 0)\\
   & = \left|m^x_{01}\right|^2\left|m^x_{12}\right|^2.
  \end{split}
\end{equation}
Similarly, $\left|\braket{\psi_1|(M^x)^2|0}\right|^2 \approx \left|m^x_{02}\right|^2\left|m^x_{12}\right|^2$.
Thus, ignoring the contributions from higher excited states, the peak intensities at $(\omega_t, \omega_\tau) = (\omega_2 - \omega_1, \pm \omega_1)$ and $(\omega_2-\omega_1, \pm \omega_2)$ serve as a measure of the transition amplitudes $\left|\braket{\psi_2|(M^x)^2|0}\right|^2$ and $\left|\braket{\psi_1|(M^x)^2|0}\right|^2$ respectively.

\section{Hybridization of two dipolar magnon modes}

\begin{figure}
  \centering
  \includegraphics[width=0.50\columnwidth]{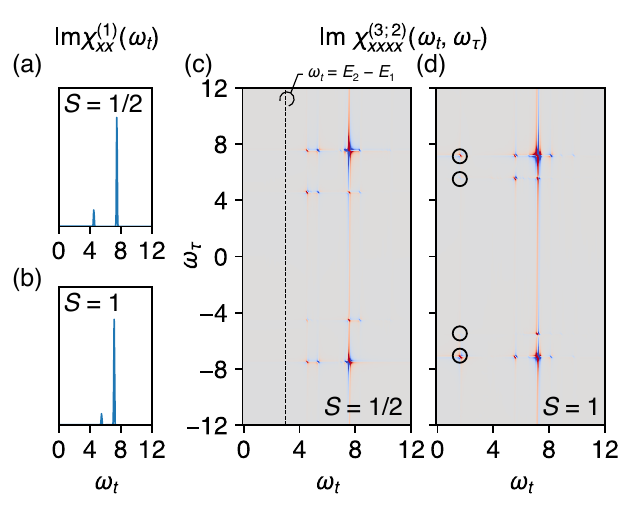}
  \caption{{\bf Signature of hybridization}.
  (a,c) Linear response $\chi^{(1)}$ and 2DCS $\chi^{(3;2)}$ of the spin-1/2 chain with staggered magnetic field [Eq.~(\ref{eq:spin1/2_staggered_field})] with $h_\text{e} = 6J$, $h_\text{o} = 8J$, $h_y = 0.4$, and $L=16$.
  (b,d) The spin-1 FM Heisenberg chain. The same parameters as the Fig.~\ref{fig:hybridized} are used.
  }
  \label{fig:hybridized_comp}
\end{figure}

We consider the spin-1/2 XXZ chain used in the main text but with a staggered tilted field
\begin{equation}
  H = J\sum_i  [\gamma(S_i^xS_{i+1}^x + S_i^yS_{i+1}^y) + S_i^zS_{i+1}^z]
  - h_\text{e} \sum_{i:\text{even}}S_i^z -  h_\text{o}\sum_{i:\text{odd}} S_i^z - h_y \sum_i S_i^y,
  \label{eq:spin1/2_staggered_field}
\end{equation}
with an AFM $J = 1$, $\gamma = 1.1$, $h_\text{e} = 6$, $h_\text{o} = 8$, and $h_y = 0.4$. The $h_y$ field breaks the $U(1)$ spin rotational symmetry of the model and the staggered $h_z$ field produces two dipolar magnons with distinct energies.
Thus, in linear response, we see two peaks, as shown in Fig.~\ref{fig:hybridized_comp}(a). Due to the finite exchange coupling $J$, the two magnon modes hybridize with one another. This means that, in the 2DCS spectrum, we observe off-diagonal peaks at $(\omega_t, \omega_\tau) = (\pm \omega_1, \pm \omega_2), (\mp \omega_1, \pm \omega_2)$ [Fig.~\ref{fig:hybridized_comp}(c)], a signature of the coherent coupling between the two modes. However, unlike the hybridization between the SIBS and dipolar magnon discussed in the main text, there are no additional peaks at $\omega_t = \omega_2 - \omega_1$ as a consequence of the vanishing quadrupolar character $\braket{\psi_i|(M^x)^2|0}$ for the two dipolar magnon modes of the spin-$1/2$ model. The 2DCS response shown in Fig.~\ref{fig:hybridized_comp}(c) and (d) for the spin-$1/2$ and spin-$1$ models respectively are thus clearly distinct, with the spin-$1/2$ model lacking any signals along the $\omega_t = \omega_2 - \omega_1$ line. 

Note that, for the particular spin-$1/2$ case considered here, the possibilities for intermediate $\ket{Q}$ states include three kinds of 2-magnon states -- created by flipping two spins at distant sites, specifically even-odd, odd-odd, and even-even configurations -- as well as an exchange-bound state created by flipping two spins at adjacent sites. These states give rise to additional peaks at various frequencies that are distinct from the diagonal and off-diagonal cross peaks discussed above, and are absent from the spin-$1$ model.

\section{Generalized spin wave theory}
We present a generalized spin wave theory (GSWT) calculation of the 2DCS spectrum, following the discussion in Ref.~\cite{Bai2021}. For clarity, we begin the discussion with the general case and then apply it to the FM ordered case in \ref{sec:GSWT_FM}, where the calculations can be considerably simplified. In \ref{sec:GSWT_2DCS}, we derive the transition amplitude $A_{PQR}$, which we used to explain features of the 2DCS spectrum in the main text.

The Hamiltonian of the spin-1 model with easy-axis single-ion anisotropy can be written as:
\begin{equation}
  \mathcal{H} = -J\sum_{\braket{i,j}} \vec{S}_i \cdot \vec{S}_{j} - \frac{D}{\sqrt{3}}\sum_i Q_i^{3z^2-r^2}.
  \label{eq:spin1_Hamiltonian_SWT}
\end{equation}
First, we introduce the SU(3) Schwinger bosons $(b_{i,1}, b_{i,0}, b_{i,-1})$ at each site $i$, with the constraint $\sum_{m} b^{\dagger}_{i,m}b_{i,m} = M$, where $M=1$. There is a correspondence between  $\ket{S^z_i = m}$ and $b^{\dagger}_{i,m}\ket{\emptyset}$, where $\ket{\emptyset}$ is the vacuum of Schwinger bosons. The spin operators can be represented as $S^{\mu}_i = \phi^{\dagger}_i L^{\mu} \phi_i$ $(\mu = x,y,z)$, where $\phi_i = (b_{i,1}, b_{i,0}, b_{i,-1})^{\text{T}}$ and $L^{\mu}$ are spin-1 matrices. Then, we introduce an SU(3) rotation $U_i$ acting on $\phi_i$, which defines another bosonic operators $\beta_{i,m} = \sum_{n} (U^\dagger_i)_{mn} b_{i,n}$. $U_i$ is chosen such that the mean-field energy $\braket{\psi|H|\psi}$ is minimized, where $\ket{\psi} = \prod_i \beta^\dagger_{i,+1}\ket{\emptyset} = \prod_i \ket{\psi_i}$; the ground state is described by the condensation of the $\beta_{i,+1}$ boson. In particular, when $U_i \in$ SU(2), the local spin has a local quantization axis $\vec{n}_i = \braket{\psi_i| \vec{S} |\psi_i}$. In that case, the $\beta_{i,0}$ and $\beta_{i,-1}$ bosons are local dipolar and quadrupolar fluctuations around the mean-field ground state, respectively.

Given that the ground state is the condensation of $\beta_{i,+1}$ bosons, where $\braket{\beta_{i,+1}} = \braket{\beta^\dagger_{i,+1}} \approx \sqrt{M}$, the spin wave Hamiltonian is now expressed using the $\beta_{i,0}$ and $\beta_{i,-1}$ bosons. Assuming $\braket{\beta^\dagger_{i,0}\beta_{i,0}} = \braket{\beta^\dagger_{i,-1}\beta_{i,-1}} \ll M$, $\beta_{i,+1}$ can be expanded as:
\begin{equation}
  \begin{split}
  \beta_{i,+1} = \beta^{\dagger}_{i,+1} & = \sqrt{M - \beta^{\dagger}_{i,0}\beta_{i,0} - \beta^{\dagger}_{i,-1}\beta_{i,-1}}\\
  & \simeq \sqrt{M}\left(1 - \frac{1}{2M}\beta^{\dagger}_{i,0}\beta_{i,0} - \frac{1}{2M}\beta^{\dagger}_{i,-1}\beta_{i,-1} + \mathcal{O}\left(\frac{1}{M^2}\right)\right).
  \end{split}
\end{equation}
Then, the spin opearators and quadrupolar operators can be expanded as
\begin{equation}
  S^{\mu}_{i} = M \mathcal{S}^{\mu}_{c}(i) + \sqrt{M} \sum_{m\neq1} \left( \mathcal{S}^{\mu}_{1m}(i) \beta_{i,m} + h.c.\right) + \sum_{m,n\neq1} \mathcal{S}^{\mu}_{mn}(i) \beta^{\dagger}_{i,m}\beta_{i,n} + \mathcal{O}\left(\frac{1}{\sqrt{M}}\right),
  \label{eq:semi_classical_S}
\end{equation}
\begin{equation}
  Q^{3z^2-r^2}_{i} = M \mathcal{Q}^{3z^2 - r^2}_{c}(i) + \sqrt{M} \sum_{m\neq1} \left(\mathcal{Q}^{3z^2 - r^2}_{1m}(i)\beta_{i,m} + h.c.\right) + \sum_{m,n\neq1} \mathcal{Q}^{3z^2-r^2}_{mn}(i)\beta^{\dagger}_{i,m}\beta_{i,n} + \mathcal{O}\left(\frac{1}{\sqrt{M}}\right),
  \label{eq:semi_classical_Q}
\end{equation}
where
\begin{equation}
  \begin{split}
    &\mathcal{S}^{\mu}_c(i) = \left( U^{\dagger}_i L^{\mu} U_i \right)_{11},\,
    \mathcal{S}^{\mu}_{1m}(i) = \left( U^{\dagger}_i L^{\mu} U_i \right)_{1m},\,
    \mathcal{S}^{\mu}_{mn}(i) = \left( U^{\dagger}_i L^{\mu} U_i \right)_{mn} - \left( U^{\dagger}_i L^{\mu} U_i \right)_{11} \delta_{mn}, \\
    &\mathcal{Q}^{3z^2-r^2}_c(i) = \left( U^{\dagger}_i Q^{3z^2-r^2} U_i \right)_{11},\,
    \mathcal{Q}^{3z^2-r^2}_{1m}(i) = \left( U^{\dagger}_i Q^{3z^2-r^2} U_i \right)_{1m},\\
    &\mathcal{Q}^{3z^2-r^2}_{mn}(i)  = \left( U^{\dagger}_i Q^{3z^2-r^2} U_i \right)_{mn} - \left( U^{\dagger}_i Q^{3z^2-r^2} U_i \right)_{11} \delta_{mn}.
  \end{split}
  \label{eq:coefficients}
\end{equation}
By inserting Eq.~(\ref{eq:semi_classical_S}, \ref{eq:semi_classical_Q}) into Eq.~(\ref{eq:spin1_Hamiltonian_SWT}), we obtain the spin wave Hamiltonian. We will do this in the following for the somewhat simpler case of a ferromagnetic ground state on a lattice with a single-site unit cell relevant for our study.

\subsection{Ferromagnetic order}
\label{sec:GSWT_FM}
In the FM ordered case, the fully polarized state along the $z$-axis minimizes the mean field energy, i.e. $U_i$ is the identity matrix, and therefore $(\beta_{i,+1}, \beta_{i, 0}, \beta_{i,-1}) = (b_{i,+1}, b_{i,0}, b_{i,-1})$ holds. Eq.~(\ref{eq:coefficients}) is now given by
\begin{equation}
  \begin{split}
    \mathcal{S}^{z}_c(i) &= \frac{1}{\sqrt{3}}\mathcal{Q}^{3z^2-r^2}_c(i) = 1,\, \mathcal{S}^{x}_c(i) = \mathcal{S}^{y}_c(i) = \mathcal{S}^{\mu}_{1m}(i) = \mathcal{Q}^{3z^2-r^2}_{1m}(i) = 0,\\
    \mathcal{S}^{\mu}_{mn}(i) &= \left( L^{\mu}  \right)_{mn} - \mathcal{S}_c^\mu(i)\delta_{mn},\, \mathcal{Q}^{3z^2-r^2}_{mn}(i) = \left( Q^{3z-r^2} \right)_{mn} -  \sqrt{3}\delta_{mn},
  \end{split}
\end{equation}
and Eqs. (\ref{eq:semi_classical_S}, \ref{eq:semi_classical_Q}) become
\begin{equation}
  \begin{split}
    S_i^z &\simeq M  - \beta^\dagger_{i,0}\beta_{i,0} - 2\beta^{\dagger}_{i,-1}\beta_{i,-1} + \mathcal{O}\left(\frac{1}{\sqrt{M}}\right), \\
    S_i^x &\simeq \sqrt{\frac{M}{2}}\left( \beta_{i,0} + \beta^\dagger_{i,0} \right) +
    \frac{1}{\sqrt{2}}\left(\beta^\dagger_{i,0}\beta_{i,-1} + \beta^\dagger_{i,-1}\beta_{i,0}\right) + \mathcal{O}\left(\frac{1}{\sqrt{M}}\right), \\
    S_i^y &\simeq -i\sqrt{\frac{M}{2}}\left( \beta_{i,0} - \beta^\dagger_{i,0} \right) -
    \frac{i}{\sqrt{2}}\left(\beta^\dagger_{i,0}\beta_{i,-1} - \beta^\dagger_{i,-1}\beta_{i,0}\right) + \mathcal{O}\left(\frac{1}{\sqrt{M}}\right), \\
    \frac{1}{\sqrt{3}}Q^{3z^2-r^2}_i &\simeq M - \beta^\dagger_{i,0}\beta_{i,0} + \mathcal{O} \left(\frac{1}{\sqrt{M}}\right).
  \end{split}
  \label{eq:expansion_operators}
\end{equation}
Inserting these into Eq.~(\ref{eq:spin1_Hamiltonian_SWT}), we obtain
\begin{equation}
  \begin{split}
    \mathcal{H}_{\text{1D}} &\simeq \mathcal{E}^{(0)} + \mathcal{H}^{(2)} + \mathcal{O}(M^0) \\
        \mathcal{H}^{(2)} &= JM \sum_{\avg{i,j}} \left(\beta^\dagger_{i,0}\beta_{i,0} + 2\beta^\dagger_{i,-1}\beta_{i,-1} -  \beta_{i,0}\beta^\dagger_{j,0} - \beta^\dagger_{i,0}\beta_{j,0} \right) + D\sum_i \beta^\dagger_{i,0}\beta_{i,0}.
  \end{split}
\end{equation}
Setting $M=1$, and using $\beta_{k,m} = (1/\sqrt{N})\sum_i e^{i\vec{k}\cdot\vec{x}_i}\beta_{i,m}$, then $\mathcal{H}^{(2)}$ can be diagonalized as
\begin{equation}
  \mathcal{H}^{(2)} = \sum_\vec{k} \left(zJ + D - J\sum_\delta e^{i\vec{k}\cdot\vec{e}_\delta}\right)\beta^\dagger_{\vec{k},0}\beta_{\vec{k},0} + \sum_\vec{k} 2zJ\beta^\dagger_{\vec{k},-1}\beta_{\vec{k},-1},
\end{equation}
where $z$ is the coordination number, and $\vec{e}_\delta$ is the unit vector along the $\delta$-th bond direction.
In particular, at $\vec{k} = 0$, the dipolar excitation energy is $D$, consistent with the ED result from the main text. On the other hand, the energy of the SIBS does not depend on $D$, meaning that linear GSWT fails to capture the energy of the SIBS at small $D/J$. 

\subsection{Calculation of susceptibilities for the FM case}
\label{sec:GSWT_2DCS}
Here, we move to the calculation of 2DCS spectra within linear GSWT. We consider a subspace $\mathcal{S}_{0,1,2}$ spanned by the basis $\{\ket{\text{vac}},\ket{i},\ket{i,j}\}$ with $\ket{i} = \beta^\dagger_i\ket{\text{vac}}$ and $\ket{i \leq j} = \zeta_{i,j}\beta^\dagger_i\beta^\dagger_j\ket{\text{vac}}$, where $i$ stands for the dictionary index of $(k_i, \sigma_i)$, $\ket{\text{vac}}$ refers to the vacuum of the $\beta_{k_i,\sigma_i\neq1}$ bosons and $\zeta_{i\neq j} = 1$ and $\zeta_{i = j} = 1/\sqrt{2!}$ are normalization factors. Note that we include the $\ket{\text{vac}}$ state in the basis, which is not necessary to include in the original GSWT formulation which focused on linear response~\cite{Bai2023}. The reason we need it here is that, with multiple pulses in the 2DCS setup, we may have the ground state as an intermediate state. Introducing the projector $P_{0,1,2}$ to the subspace $\mathcal{S}_{0,1,2}$, the restricted Hamiltonian is obtained by the projection
\begin{equation}
\mathcal{H}_{\text{eff}} = P_{0,1,2}\mathcal{H}P_{0,1,2} = \begin{pmatrix}
0 & 0 & 0 \\
0 & H_{11} & H_{12}\\
0 & \text{h.c.} & H_{22}
\end{pmatrix}.
\end{equation}
In linear GSWT, the effective Hamiltonian $\mathcal{H}^{(2)}_{\text{eff}}$ is already diagonal in the subspace $\mathcal{S}_{0,1,2}$, and we have
\begin{equation}
  H^{i, j\leq k}_{12} = 0,\, H^{ij}_{11} = \delta_{ij}\epsilon_i,\, H^{i\leq j, k \leq l}_{22} = \delta_{ik}\delta_{jl}(\epsilon_i + \epsilon_j).
\end{equation}
Now, we expand the magnetization operator $M^x$ up to $\mathcal{O}(1)$. In particular, for the FM case, we have
\begin{equation}
  M^x = \sum_i\mathcal{S}_i^x \simeq \sqrt{\frac{MN}{2}}\left(\beta_{\vec{k}=\vec{0},0} + \beta^\dagger_{\vec{k} = \vec{0},0}\right) + \frac{1}{\sqrt{2}}\sum_{\vec{k}}\left(\beta^\dagger_{\vec{k},0}\beta_{\vec{k},-1} + \beta_{\vec{k},0}\beta^\dagger_{\vec{k},-1}\right).
  \label{eq:Mx_expansion}
\end{equation}
The only state in which $M^x\ket{\text{vac}}$ can have a finite overlap is the 1-magnon state $\beta^\dagger_{\vec{k}=\vec{0},0}\ket{\text{vac}}$, meaning that the intermediate states $\ket{P}$ and $\ket{R}$ are always $\beta^\dagger_{\vec{k}=\vec{0},0}\ket{\text{vac}}$ with $E_P = E_R = D$. On the other hand, the intermediate state $\ket{Q}$ can be one of $\ket{\text{vac}}$, $\beta^\dagger_{\vec{k}=\vec{0},-1}\ket{\text{vac}}$, and $(1/\sqrt{2})(\beta^\dagger_{\vec{k}=\vec{0},0})^2\ket{\text{vac}}$, with energies $E_Q =0$, $2zJ$, and $2D$, respectively. The matrix elements required for the calculation of the transition amplitude $A_{PQR}$ can be calculated as
\begin{equation}
    \braket{\text{vac}|M^x|P}  = \braket{\text{vac}|M^x|R}  =  \sqrt{\frac{N}{2}},\,
    \braket{P|M^x|Q} = \braket{R|M^x|Q} = \begin{dcases}
      \sqrt{\frac{N}{2}} &  \ket{Q} = \ket{\text{vac}}\\
      \frac{1}{\sqrt{2}} &  \ket{Q} = \beta^\dagger_{\vec{k} = \vec{0},-1}\ket{\text{vac}}\\
      \sqrt{N} &   \ket{Q} = \frac{1}{\sqrt{2}}(\beta^\dagger_{\vec{k} = \vec{0},0})^2\ket{\text{vac}}\\
    \end{dcases}.
\end{equation}
Therefore, the overall transition probability $A_{PQR} = m_{0P}m_{PQ}m_{QR}m_{R0}$ is $N^2/4$ when $\ket{Q}=\ket{\text{vac}}$, and $N^2/2$ when $\ket{Q}=(1/\sqrt{2})(\beta^\dagger_{\vec{k} = \vec{0},0})^2\ket{\text{vac}}$. Combined with the fact that the 2-magnon state has exactly twice the energy of the 1-magnon state, one can show that the contribution from these two processes exactly cancel out~\cite{Watanabe2024}. The contribution from the other process, with $\ket{Q} = \beta^\dagger_{\vec{k}=\vec{0},-1}\ket{\text{vac}}$, only gives the SIBS peaks, but not the PP and 2Q peaks.

The failure of linear GSWT to recover the PP and 2Q peaks is partly due to overlooking the inability to generate two excitations simultaneously at a single site, which we expect nonlinear GSWT to capture. 
Accounting for quasiparticle interactions changes the energy and matrix element of the process where $\ket{Q}$ is the 2-magnon state, impairing perfect cancellation and leading to the appearance of the missing peaks. Here, instead of considering the nonlinear corrections to the GSWT Hamiltonian itself, we use the following unexpanded $S^x_i$ for the magnetization operator $M^x$ to calculate the transition probability
\begin{equation}
  \mathcal{S}_i^x = \frac{1}{\sqrt{2}}\left(\sqrt{1 - \beta_{i,0}^\dagger\beta_{i,0} - \beta_{i,-1}^\dagger\beta_{i,-1}}\beta_{i,0} + \beta_{i,0}^\dagger\sqrt{1 - \beta_{i,0}^\dagger\beta_{i,0} - \beta_{i,-1}^\dagger\beta_{i,-1}} + 
  \beta_{i,-1}^\dagger\beta_{i,0} + \beta_{i,0}^\dagger\beta_{i,-1}\right).
\end{equation}
With this modification, we have an additional process in which the intermediate $\ket{Q}$ state is $\beta^\dagger_{\vec{k},0}\beta^\dagger_{-\vec{k},0}\ket{\text{vac}}$ ($\vec{k} \neq \vec{0}$) with energy $E_Q = 2zJ + 2D - 2zJ\sum_\delta\cos(\vec{k}\cdot\vec{e}_\delta)$. 
The relevant matrix elements are given by
\begin{equation}
    \braket{\text{vac}|M^x|P}  = \braket{\text{vac}|M^x|R}  =  \sqrt{\frac{N}{2}},\,
    \braket{P|M^x|Q} = \braket{R|M^x|Q} = \begin{dcases}
      \sqrt{\frac{N}{2}} &  \ket{Q} = \ket{\text{vac}}\\
      \frac{1}{\sqrt{2}} &  \ket{Q} = \beta^\dagger_{\vec{k} = \vec{0},-1}\ket{\text{vac}}\\
      \frac{N-1}{\sqrt{N}} &   \ket{Q} = \frac{1}{\sqrt{2}}(\beta^\dagger_{\vec{k} = \vec{0},0})^2\ket{\text{vac}}\\
      -\sqrt{\frac{2}{N}} &   \ket{Q} = \beta^\dagger_{\vec{k},0}\beta^\dagger_{-\vec{k},0}\ket{\text{vac}}
    \end{dcases},
\end{equation}
Note that, in the current FM case, $M_x$ can only create at most one additional quasiparticle, and the number of quasiparticles is conserved in the bosonic Hamiltonian, so limiting the Hilbert space to the subspace $\mathcal{S}_{0,1,2}$ itself is exact as long as zero temperature 2DCS is considered.
The approximation used here is the expansion of the square root [Eq.~(\ref{eq:expansion_operators})].

\end{document}